\documentclass[a4paper,accepted=2023-08-07]{quantumarticle}

\pdfoutput=1 

\usepackage[utf8]{inputenc}
\usepackage[english]{babel}
\usepackage[T1]{fontenc}
\usepackage{amsmath}

\usepackage{amssymb}
\usepackage{amsfonts}
\usepackage{amsthm}

\usepackage{bm}
\usepackage{csquotes}
\usepackage[normalem]{ulem}
\usepackage[dvipsnames]{xcolor}

\PassOptionsToPackage{hyphens}{url}
\usepackage[breaklinks,colorlinks=true]{hyperref}
\usepackage[capitalise]{cleveref}
\usepackage{graphicx}

\usepackage{physics}
\usepackage{mathtools}
\usepackage{xfrac}

\usepackage{enumitem}
\usepackage{placeins}
\usepackage{float}
\usepackage[caption=false]{subfig}

\newtheorem{theorem}{Theorem}
\newtheorem{corollary}{Corollary}
\newtheorem{definition}{Definition}

\DeclareMathOperator*{\argmin}{argmin}

\newcommand{\innerprod}[2]{{ \langle {#1}, {#2} \rangle  }}

\begin{document}

\title{Learning ground states of gapped quantum Hamiltonians with Kernel Methods}
\author[1,2]{Clemens Giuliani}
\author[1,2]{Filippo Vicentini}
\author[1,2,3]{Riccardo Rossi}
\author[1,2]{Giuseppe Carleo}
\affil[1]{Institute of Physics, \'{E}cole Polytechnique F\'{e}d\'{e}rale de Lausanne (EPFL), CH-1015 Lausanne, Switzerland}
\affil[2]{Center for Quantum Science and Engineering, \'{E}cole Polytechnique F\'{e}d\'{e}rale de Lausanne (EPFL), CH-1015 Lausanne, Switzerland}
\affil[3]{Sorbonne Universit\'e, CNRS, Laboratoire de Physique Th\'eorique de la Mati\`ere Condens\'ee, LPTMC, F-75005 Paris, France}

\begin{abstract}
Neural network approaches to approximate the ground state of quantum hamiltonians require the numerical solution of a highly nonlinear optimization problem.
We introduce a statistical learning approach that makes the optimization trivial by using kernel methods.
Our scheme is an approximate realization of the power method, where supervised learning is used to learn the next step of the power iteration.
We show that the ground state properties of arbitrary gapped quantum hamiltonians can be reached with polynomial resources under the assumption that the supervised learning is efficient.
Using kernel ridge regression, we provide numerical evidence that the learning assumption is verified by applying our scheme to find the ground states of several prototypical interacting many-body quantum systems, both in one and two dimensions, showing the flexibility of our approach. 
\end{abstract}

\maketitle

\section{Introduction}
The exact simulation of quantum many-body systems on a classical computer requires computational resources that grow exponentially with the number of degrees of freedom.
However, to address scientifically relevant problems such as strongly-correlated materials~\cite{lee_wen_rmp_2006_mott_insulatir_high_temp_supercond, kopp_nature_physics_2005_criticality_in_correlated_quantum_matter} or quantum chemistry~\cite{Dral2020JPCL,McArdle2020RMP}, it is necessary to study large systems.
Over the years, a variety of numerical methods have been proposed that exploit the specific structure of the system at hand and  resort to approximation schemes to lower the computational cost.
For example, tensor networks~\cite{white_prl_1992_dmrg}
can efficiently encode one-dimensional systems, but  they face challenges in higher dimensions~\cite{verstraete_cirac_arxiv_2004_peps}.
Quantum Monte-Carlo methods~\cite{ceperley_science_1986_quantum_monte_carlo, becca_sorella_book_2017} can give accurate results for stoquastic Hamiltonians~\cite{bravyi_arxiv_2006_stoquastic}, but is in general plagued by the so-called \emph{sign-problem}~\cite{loh_prb_1990_sign_problem, troyer_prl_2005_sign_problem}.
Finally, traditional variational methods~\cite{jastrow_pr_1955, bardeen_cooper_shrieffer_pr_1957_bcs, sorella_prl_1998_green_function_monte_carlo_sr, becca_sorella_book_2017} require that the ground- or time-evolving state be well approximated by a physically-inspired parameterized function~\cite{YokoyamaShiba1987,GrossJoyntRice1987,Gross1988}.

Recently, {\it data-driven} approaches for compressing the wave function based on neural networks~\cite{carrasquilla_melko_nature_2017_ml_quantum_phases_of_matter, torlai_carleo_nature_physics_2018_quantum_state_tomography} and kernel methods~\cite{glielmo_booth_prx_2020_gaussian_process_states, rath_booth_jcp_2020_bayesian_opt_quantum_state,luo_arxiv_2021_ntk} have been proposed.
However, data-driven approaches require knowledge of the exact wave function, either from experiments or from numerically-exact calculations.
In contrast, the variational principle for ground-state calculations provides a {\it principle-driven} approach to wave function optimization problem, and it has lead to the proposal of Neural(-network) Quantum States (NQS)~\cite{carleo_troyer_science_2017} and Gaussian process states~\cite{rath_booth_prr_2022_gaussian_process_states,Rath2023BoothArxivGPSElectrons}. 
The NQS approach produced state-of-the-art ground-state results on a variety of systems such as the $J_1$-$J_2$ spin model~\cite{nomura_prx_2021_j1j2,roth_arxiv_2021_gcnn_nqs,astrakhantsev_carleo_neupert_prx_2021_pyrochlore}, atomic nuclei~\cite{lovato_carleo_prr_2022_nqs_atomic_nuclei}, and molecules~\cite{zhao_stokes_arxiv_2022_nqs_quantum_chemistry}.
While neural-networks are universal function approximators and can, in principle, represent any wave-function, in practice the variational energy optimization of NQS is a non-trivial task~\cite{westerhout_nature_comm_2020_nqs_generalization_frustration_learn_sign, szabo_prr_2020_nqs_sign_problem}. 
Ref.~\cite{kochkov_clark_arxiv_2018_supervised_wavefunction_optimization,jonsson_baur_arxiv_2018_nqs_qc_sim,atanasova_naturecomm_2023_stoch_repr_qs} proposed schemes which solve a series of simpler supervised-learning tasks instead.
When used to solve the ground-state problem with a first-order approximation of imaginary time evolution with a large time step, equivalent to the power method, we refer to this approach as the Self-Learning Power Method (SLPM).
This supervised approach does not immediately solve the optimization hardness of NQS, as there is still a non-trivial optimization problem to solve at every step of the procedure.

Kernel methods are a popular class of machine learning methods for supervised learning tasks~\cite{shawe_book_2004_kernel_methods, smola_book_2008_kernel_methods_in_ml}.
They map the input data to a high-dimensional space by a non-linear transformation, with the goal of making the input data correlations approximately linear.
The similarity between the input data is encoded by the kernel, whose choice is problem-dependent.
When compared to neural-network approaches, kernel methods have the crucial advantage that the solution of certain optimization problems can be obtained by solving a linear system of equations.

In this article we combine the SLPM with kernel methods, rendering the optimization problem at each step of the power method straightforward.
We prove the convergence of SLPM methods to the ground state of gapped quantum Hamiltonians under a learning-efficiency assumption by generalizing previous results of Ref.~\cite{hard_neurips_2014_noisy_power_method}.
Considering the SLPM with kernel ridge regression, we numerically verify the learning-efficiency assumption for small quantum systems.
For larger systems, we estimate the ground-state energy directly and find a favorable system-size scaling.

The article is organized as follows: in \cref{sec:preliminaries}, we recall the power method and the basics of supervised learning, setting the notation we use throughout the text.
In \Cref{sec:self_learning_power_method}, we introduce the SLPM, with \cref{sec:convergence} containing an in-depth theoretical analysis of its convergence properties, which is the first major result of our work.
Then, after briefly recalling Kernel Ridge Regression in \cref{sec:kernel_ridge_regression}, we discuss our particular choice of kernel and numerical implementation of the SLPM in \cref{sec:slpm-with-krr,sec:kernel_choice}.
Finally, in \cref{sec:results}, we provide comprehensive numerical results obtained on the transverse-field Ising (TFI) and antiferromagnetic Heisenberg (AFH) models in one and two dimensions, concluding with a discussion in \cref{sec:discussion}.

\section{Preliminaries}
\label{sec:preliminaries}
First, we briefly introduce our notation and recap some well-known concepts regarding the Power Method (PM), in  \cref{sec:pwr}. We then briefly overview supervised learning in \cref{sec:superv}.

Let $\hat H$ be a hamiltonian of a quantum system.
We denote its normalized eigenstates by $\ket{\Upsilon_k}$, and we order them with respect to their corresponding eigenvalues $E_k$, such that  $E_0 \le E_1 \le \cdots \le E_{\mathrm{max}}$.
We wish to determine the ground state $\ket{\Upsilon_0}$ and its energy $E_0$.
The gap of $\hat H$ is defined as $\delta = E_1 -E_0$, and we say that the Hamiltonian is gapped if $\delta > 0$.
We remark that the method remains efficient for Hamiltonians that are gapless in the thermodynamic limit, as long as the gap closes polynomially with the inverse system size.

\subsection{Power Method}
\label{sec:pwr}
The PM is a procedure to find the dominant eigenvector\footnote{The dominant eigenvector is the eigenvector with the largest eigenvalue by magnitude.} of a matrix.
Following the notation of Ref.~\cite{becca_sorella_book_2017}, we consider a gapped hamiltonian $\hat H$ and a constant $\Lambda\in\mathbb{R}$.
The PM relies on the repeated application of the shifted Hamiltonian $\Lambda -\hat H$  to a trial state $\ket*{\Phi^{(0)}}$.
The state obtained at the $(n+1)-$th step is, therefore,
\begin{equation}
    \label{eq:pwrstep}
    \ket*{\Phi^{(n+1)}} = (\Lambda - \hat H) \ket*{\Phi^{(n)}}.
\end{equation}
To make the ground state the dominant eigenvector, one must take $\Lambda > \frac{E_0 + E_{\mathrm{max}}}{2}$.
Starting from a $\ket*{\Phi^{(0)}}$ with non-zero overlap with the true ground state  $\ket*{\Upsilon_0}$, we have that $\lim_{n \rightarrow \infty} \ket*{\Phi^{(n)}} \propto \ket*{\Upsilon_0}$,
as the infidelity with the ground state decreases exponentially with $(\frac{\Lambda -E_1}{\Lambda -E_0})^n$ as long as  $\Lambda \geq \frac{E_1 + E_{\mathrm{max}}}{2}$ and with $(\frac{E_{\mathrm{max}}-\Lambda}{\Lambda -E_0})^n$ otherwise.
We note that, in the case of a degenerate ground state, the PM converges to a linear combination of the degenerate eigenstates, dependent on their overlap with the initial state $\ket*{\Upsilon_0}$.

The PM is widely adopted in exact diagonalization studies, where one works with vectors storing the wave-function amplitude $\braket*{x}{\Phi^{(n)}}$ for all the basis states $x$.
This approach requires exponential resources to store the vector encoding the wave function amplitudes in a chosen basis.

\subsection{Supervised Learning}
\label{sec:superv}
Suppose we are given a set of observations $\mathcal D = \{(x_i, y_i)\}_{i=1}^{N_s}$
of an unknown function $f^0 : \mathcal X \to \mathbb R$ where $\mathcal X \subseteq \mathbb R^n$ is the input space, $x_i \in \mathcal X$ are samples and $y_i = f^0(x_i) \in \mathbb R$ are the corresponding function values, also known as \textit{labels}.
The task of supervised learning is to find the optimal function $f^\star$ in some suitable space of Ansatz functions $\mathcal H$ which best describes the observations.
This is done by minimizing a so-called \textit{loss function} $\mathcal L$, which quantifies the distance between the predictions of each Ansatz function and the observations.
The corresponding optimization problem is given by
\begin{equation}
    \label{eq:supervised_opt_problem}
    f^\star \in \argmin_{f \in \mathcal H} \mathcal{L}(f, \mathcal D).
\end{equation}

Notably, supervised learning can be done with artificial neural networks, a specific instance of highly expressive parameterized maps that can typically approximate complex unknown functions with high accuracy.
After fixing the architecture, the optimization problem \cref{eq:supervised_opt_problem} is solved by finding the optimal parameters, for example, by using gradient-based optimization methods.
For a more complete overview of supervised learning, we refer the reader to one of the standard textbooks in the literature, such as Ref.~\cite{russell_book_1995_2020_artificial_intelligence}.

\section{Self-Learning Power Method}
\label{sec:self_learning_power_method}

\begin{figure*}[htb]
    \includegraphics[width=\textwidth]{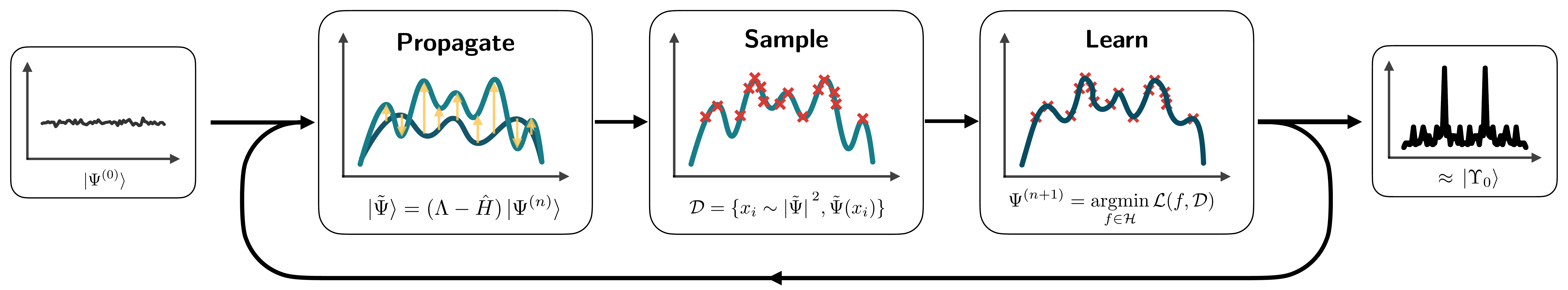}
    \caption{{\it Sketch of the Self-Learning Power Method}.
Starting from an initial state $\Psi^{(0)}$, we propagate the state $\Psi^{(n)}$ at step $(n+1)$ to $\tilde \Psi$ by applying $\Lambda -\hat H$. Configurations $x_i$ are sampled from $\abs*{\tilde \Psi(x)}^2$, paired with wave-function amplitudes $\tilde \Psi(x_i)$ to form a data set, which is then learned with supervised learning, obtaining a new state $\Psi^{(n+1)}$, approximating $\tilde \Psi$, which can again be propagated and sampled from.
The procedure is repeated until convergence to a state close to the true ground state. In this paper the learning is done with kernel ridge regression (see \cref{sec:slpm-krr-section}).}
    \label{fig:fig1}
\end{figure*}

In this section, we introduce an approximate version of the PM that has polynomial complexity (\cref{sec:SelfLearningAlgorithm}) and provide a quantitative theoretical discussion of its convergence properties (\cref{sec:convergence}).
We call this approach the Self-Learning Power Method (SLPM), which is sketched in \cref{fig:fig1}.
The SLPM encodes the wave function with an approximate representation $\ket{\Psi^{(n)}}\approx\ket{\Phi^{(n)}}$, taken from a space $\mathcal{H}$ of functions with a polynomial memory and query complexity in the computational basis\footnote{By query complexity in the computational basis we mean that computing $\braket{x}{\Psi^{(n)}} = \Psi^{(n)}(x)$ requires polynomial resources in the system size.} to bypass the exponential computational cost of the exact PM, as discussed in the previous section.
In the following, we show that the state at step $n+1$ can be computed by solving an optimization problem given the state at step $n$.

\subsection{Algorithm}
\label{sec:SelfLearningAlgorithm}

Given $\Psi^{(n)}$ the state $\Psi^{(n+1)} \in \mathcal H$ is the solution of the optimization problem
\begin{equation}
    \label{eq:slpm_opt_problem_general}
    \Psi^{(n+1)} \in \argmin_{f \in \mathcal H} \mathcal{L}(f, (\Lambda-\hat H) \Psi^{(n)}),
\end{equation}
for any similarity metric $\mathcal L$.
In this article, we treat this optimization problem in the framework of supervised learning, which we have introduced in \cref{sec:superv}, replacing the "target" state $(\Lambda-\hat H) \Psi^{(n)}$ with a data-set $\mathcal{D}^{(n+1)} = \{(x_i, y_i)\}$ where
\begin{equation}
\label{eq:data-set}
\begin{aligned}
&x_i \sim \Pi(x) = \abs{\mel*{x}{\Lambda-\hat H}{\Psi^{(n)}}}^2 \\
&y_i = \mel*{x_i}{\Lambda-\hat H}{\Psi^{(n)}}
\end{aligned}
\end{equation}
Here $\sim$ indicates that $x_i$ are sampled from the target distribution $\Pi$, which we do with Markov-chain Monte Carlo methods (see  \cref{appendix:sampling} for a discussion).
We remark that most physical Hamiltonians are sparse and therefore the elements $\mel*{x}{\Lambda-\hat H}{\Psi^{(n)}}$ can be queried efficiently if $\Psi^{(n)}$ can be queried efficiently in the computational basis\footnote{More in detail, one has to calculate $\sum_{x\prime} \mel*{x}{\Lambda - \hat H}{x^{\prime}} \braket*{x^{\prime}}{\Psi^{(n)}}$.
Physical Hamiltonians, in general, have a polynomial number of nonzero terms $\mel*{x}{\hat H}{x^{\prime}} \neq 0$, therefore the sum over $x'$ runs over a polynomial number of elements and this query is efficient. }.

In practice, when considering spin systems on a lattice with $N$ sites, the wave-function $\psi(x)$ takes as inputs the bit-strings $x \in \{-1, 1\}^N$ encoding basis states $\ket x$.
The data set contains a polynomially-large set of bit-strings $x$, sampled from the Born-probability distribution $\abs{\psi(x)}^2$, and associated with their corresponding amplitude $\psi(x)$.

We remark that if $\mathcal{H}$ spans the whole Hilbert space and if the data set contains all (exponentially-many) bit-strings, the solution to the optimization problem given by \cref{eq:slpm_opt_problem_general} would match the PM exactly.
By truncating the data-set size and considering only a subset of all possible wave functions, the solution is only approximate and therefore there is a finite difference between an exact step of the PM and the approximate procedure.
We quantify this difference with the \textit{step infidelity}, which we define as
\begin{definition}[\textbf{Step Infidelity}]
\label{def:stepfid}
Let $\ket{\Psi^{(n+1)}}$ be the state after step $n$ of the noisy power method.
We define the step infidelity as
\begin{equation}
\label{eq:stepfid_def}
I^{(n)} \coloneqq 1-\mathcal{F}\left(\Psi^{(n+1)}, (\Lambda-\hat{H})\Psi^{(n)}\right)\\
\end{equation}
where $\mathcal{F}(\psi, \phi) = \frac{\abs{\braket{\psi}{\phi}}^2}{\norm{\phi}^2 \norm{\psi}^2}$ is the fidelity between two states.
\end{definition}

\subsection{Discussion of convergence properties}
\label{sec:convergence}

The SLPM is approximating the propagation of the state
\begin{equation}
\ket*{\Psi^{(n+1)}} \approx (\Lambda -\hat H) \ket*{\Psi^{(n)}},
\end{equation}
which is instead exact when using the standard PM.
It is well known (see \cref{sec:pwr}) that the PM converges exponentially fast to the dominant eigenstate. In this subsection, we discuss how the noise introduced by a non-zero step-infidelity affects the convergence.
To derive quantitative bounds, we prove that if this noise is small enough, it does not significantly hinder the convergence to the ground state as the relative error of the energy is bounded (as we will see in \cref{eq:rel_err_bound}).
The discussion is based on Ref.~\cite{hard_neurips_2014_noisy_power_method}, but has been adapted to the language of computational physics.

\subsubsection*{Power method with noise}
We consider the general case where, at every step of the PM, a small noise term $\ket{\Delta}$ is added to the state.
In the setting we are interested in, this noise arises from the self-learning procedure, but we wish to keep the theoretical treatment general and accordingly we make no assumptions on the origin of the noise.
Formally, we define the power method with noise as:

\begin{definition}[\textbf{Noisy Power Method}]
\label{def:npm}
Take an initial state $\ket*{\Psi^{(0)}}$.
Step $(n+1)$ of the noisy power method is defined recursively as
\begin{equation}
    \ket*{\Psi^{(n+1)}} = \gamma^{(n)}\left[(\Lambda-\hat{H})\ket*{\Psi^{(n)}} +\ket*{\Delta^{(n)}}\right]
\end{equation}
where $\ket{\Delta^{(n)}}$ is a additive noise term, which, without loss of generality\footnote{
    In the case that $\ket*{\Delta}$ is not orthogonal to $\ket*{\tilde \Psi} \coloneqq (\Lambda -\hat H)\ket*{\Psi} $ we can replace it with $\ket*{\Delta_\perp} \coloneqq \frac{\braket{\tilde\Psi}\ket{\Delta} - \braket{\tilde\Psi}{\Delta}\ket{\tilde\Psi}}{\braket{\tilde\Psi}+\braket{\tilde\Psi}{\Delta}}$,  then we have that $\braket*{\Delta_\perp}{\tilde\Psi} = 0$ and $\ket*{\tilde\Psi} + \ket*{\Delta_\perp} \propto \ket{\tilde\Psi} + \ket{\Delta}$, and the resulting proportionality constant is absorbed into $\gamma$.}, is taken such that $\mel*{\Delta^{(n)}}{\Lambda-\hat{H}}{\Psi^{(n)}}=0$.
The factor $\gamma^{(n)}\in\mathbb{C}$ captures both a potential drift in the global phase as well as the normalization.
\end{definition}

If $I^{(n)}=0$, the noise must be zero as well, while in the general case the step-infidelity bounds the amplitude of the noise term according to
\begin{equation}
\label{eq:error_norm_stefid_bound}
\frac{\norm{\Delta^{(n)}}}{ \norm{\Psi^{(n)}}} \leq (\Lambda - E_0) \sqrt{\frac{I^{(n)}}{1-I^{(n)}}}.
\end{equation}

\begin{theorem}[Convergence of the noisy power method]
\label{thm:thm2.3}
  Let $\ket*{\Upsilon_0}$ represent the ground state of the Hamiltonian $\hat H$.
  Take $\Lambda \geq \frac{E_1 + E_{\mathrm{max}}}{2}$ and assume that the initial state $\ket*{\Psi^{(0)}}$ and noise $\ket*{\Delta^{(n)}}$ respect the conditions
  \begin{align}
    \label{eq:assumption-1}
    \frac{\abs{\braket*{\Upsilon_0}{\Delta^{(n)}}}}{\norm{\Psi^{(n)}}} &\leq \frac{\delta}{5} \frac{\abs{\braket*{\Upsilon_0}{\Psi^{(0)}}}}{\norm{\Psi^{(0)}}}\\
    \label{eq:assumption-2}
    \frac{\norm{\Delta^{(n)}}}{\norm{\Psi^{(n)}}} &\leq \frac{\delta}{5} \varepsilon,
  \end{align}
  at every step $n$ of the noisy power method for some $\varepsilon < \frac{1}{2}$.
  Then there exists a minimum number of steps $M \leq \frac{4}{1- \frac{\Lambda -E_1}{\Lambda -E_0}} \log\left(\varepsilon^{-1}\sqrt{\frac{1-\mathcal{F}(\Upsilon_0, \Psi^{(0)})}{\mathcal{F}(\Upsilon_0, \Psi^{(0)})}}\right)$
  such that for all steps $n \geq M$ we have
  $\sqrt{\frac{1-\mathcal{F}(\Upsilon_0, \Psi^{(n)})}{\mathcal{F}(\Upsilon_0, \Psi^{(n)})}} \leq \varepsilon$.
\end{theorem}
A proof adapted from Ref.~\cite{hard_neurips_2014_noisy_power_method} is given in \cref{appendix:proof_thm_2.3}.
\cref{eq:assumption-1} requires that, if the initial state $\Psi^{(0)}$ has an exponentially small overlap with the ground state (as in random initialization~\cite{hard_neurips_2014_noisy_power_method}), the noise parallel to ground state wave-function must also be exponentially small.
Instead, the assumption of \cref{eq:assumption-2} requires that the noise amplitude be smaller than $\varepsilon$.
As the final infidelity is bounded by $\varepsilon^2$ we want to choose the smallest $\varepsilon$ possible.
For a given step infidelity $I^{(n)}$ the smallest $\varepsilon$ we can guarantee using \cref{eq:error_norm_stefid_bound} is given by

\begin{equation}
\label{eq:epsilon_relation_step_infidelity}
\varepsilon^\star = \frac{5}{1-\frac{\Lambda -E_1}{\Lambda -E_0}} \max_n \sqrt{\frac{I^{(n)}}{1-I^{(n)}}}.
\end{equation}
Requiring $\varepsilon^\star<\frac{1}{2}$, it is possible to show that the step-infidelity must be sufficiently small and satisfy $I^{(n)} < \frac{1}{100}(1-\frac{\Lambda -E_1}{\Lambda -E_0})^{2}$.

When those requirements are satisfied, \cref{thm:thm2.3} states that in a number of steps $M$, logarithmic in both the initial overlap $\abs{\braket*{\Upsilon_0}{\Psi^{(0)}}}$ and in the final infidelity $\varepsilon^2$, we reach a state with at most
\begin{equation}
\label{eq:final_infidelity_bound_epsilon}
\mathcal I = 1-\mathcal{F}(\Upsilon_0, \Psi^{(M)}) \leq \frac{1-\mathcal{F}(\Upsilon_0, \Psi^{(M)})}{\mathcal{F}(\Upsilon_0, \Psi^{(M)})} \leq \varepsilon^2.
\end{equation}
This state has an accuracy on the ground-state energy given by the relative error,
\begin{equation}
\begin{aligned}
\label{eq:rel_err_bound}
\epsilon_{\mathrm{rel}} &\coloneqq \frac{\langle\hat{H}\rangle-E_0}{\abs{E_0}} \le \frac{E_{max}- E_0}{\abs{E_0}}\,  \mathcal I \\
\end{aligned}
\end{equation}

For simplicity, the theorem assumes that the noise bounds are constant throughout the run-time of the noisy power method.
While this might not be the case in practice, it is easy to generalize the result to a varying $\varepsilon$.
Doing so, one finds that the first assumption (\cref{eq:assumption-1}) is necessary to start the method while asymptotically, the bound is given only by the latter assumption (this is discussed more in detail in \cref{appendix:extra_figures}).

\subsubsection*{Self-Learning Power Method}

While the discussion of the noisy power method convergence so far is general, we now contextualize it to the case of the SLPM.
To do so we assume that the \textit{learning is efficient}, precisely defined as follows.
\begin{definition}[Efficient supervised learning]
    We say that the supervised learning is efficient if its step-infidelity is of the order of $1/{N_{S}}^\alpha$ for some $\alpha>0$, where $N_{S}$ is the size of the data-set.
\end{definition}
As a consequence of \cref{thm:thm2.3}, summing up the discussion of the convergence properties, we present the following corollary for the convergence of the SLPM:
\begin{corollary}[Convergence of the self-learning power method]
\label{corr:self_learning_power_method_convergence}
Let $\hat H$ be a gapped Hamiltonian, take $\Lambda \geq \frac{E_1 + E_{\mathrm{max}}}{2}$,
and assume that
\begin{itemize}
    \item The supervised learning is efficient, meaning that $I^{(n)} \leq \frac{A}{ {N_S}^{\alpha}} \leq \frac{1}{100} (1-\frac{\Lambda -E_1}{\Lambda -E_0})^2$ for $A, \alpha>0$.
    \item The error parallel to the ground state is bounded by $\frac{\abs{\braket*{\Upsilon_0}{\Delta^{(n)}}}}{\norm{\Psi^{(n)}}} \leq \frac{\delta}{5} \frac{\abs{\braket*{\Upsilon_0}{\Psi^{(0)}}}}{\norm{\Psi^{(0)}}}$.
\end{itemize}
Then the final infidelity $\mathcal I$ is bounded by
\begin{equation}
    \mathcal I \leq \frac{25}{\left(1-\frac{\Lambda -E_1}{\Lambda -E_0}\right)^2} \frac{A}{ {N_S}^{\alpha}}.
\end{equation}
and the error on the ground-state energy of $\hat H$ is of the order of
\begin{equation}
    \epsilon_{\mathrm{rel}} \lesssim \frac{1}{\delta^2 {N_{S}}^\alpha}.
\end{equation}
\end{corollary}
Therefore, assuming the supervised learning is efficient, it is possible to consider a polynomially large data set to compute the ground-state energy of a gapped Hamiltonian with a polynomial cost.
The same holds for gapless Hamiltonians, as long as the gap closes polynomially with the inverse system size, as we show in \cref{appendix:gappless}.

\section{The Self-learning Power Method with Kernel Ridge Regression}
\label{sec:slpm-krr-section}

There are several practical ways to implement the SLPM defined in \cref{sec:self_learning_power_method}, by solving the supervised learning problem of learning the next state (\cref{eq:slpm_opt_problem_general}) with a suitable approach.
In this section, we specialize our discussion on realizing the SLPM with a kernel method called Kernel Ridge Regression.

\subsection{Kernel Ridge Regression}
\label{sec:kernel_ridge_regression}

Given that we are discussing a kernel method we start with a brief, formal definition of the \emph{kernel}.
A positive definite kernel $k$ (named \textit{Mercer Kernel} after the author of Ref.~\cite{mercer_royal_society_1909}), is a function ${k: \mathcal X \times \mathcal X \to \mathbb R}$, where $\mathcal X \subseteq \mathbb R^n$, with the following properties:
\begin{itemize}
    \item It is symmetric: $k(x, y) = k(y,x)$
    \item For any set $\{x_1, \dots, x_n\} \subseteq \mathcal X$ the kernel matrix $K$  with entries $K_{ij} = k(x_i, x_j)$ is positive semi-definite.
\end{itemize}
It can be shown that every kernel uniquely defines a function space~\cite{aronszajn_trans_ams_1950_moore_aronszajn_thm},
the so-called \emph{Reproducing Kernel Hilbert Space} (RKHS)
\begin{equation}
\begin{aligned}
\label{eq:reproducing_kernel_hilbert_space}
    \mathcal H_k &= \{f(\cdot) = \sum_{i=1}^{\ell} w_i k(\cdot, x_i)\, |\, \ell \in \mathbb N, w_i \in \mathbb R, x_i \in \mathcal X\}
\end{aligned}
\end{equation}
where, for two functions $f,g \in \mathcal{H}_k$, $f(x) = \sum_{i=1}^{\ell} \alpha_i k(x, x_i)$, $g(x) = \sum_{j=1}^{m} \beta_j k(x, y_j)$ the inner product is given by
${
    \innerprod{f}{g}_{\mathcal{H}_k} = \sum_{i=1}^{\ell} \sum_{j=1}^{m} \alpha_i \beta_j k(x_i, y_j).
}$

The RKHS can be used as space of Ansatz functions for the supervised learning problem \cref{eq:supervised_opt_problem}.
When using the regularized least squares loss (ridge loss)
\begin{equation}
\label{eq:krr_loss}
    \mathcal L(f, \mathcal D) = \sum_i \abs{f(x_i) - y_i}^2 + \lambda\, \norm*{f}_{\mathcal{H}_k}^2,
\end{equation}
this approach is called Kernel Ridge Regression (see e.g. Ref.~\cite{shawe_book_2004_kernel_methods} for more details).
Here  $\lambda\, \norm*{f}_{\mathcal{H}_k}^2$ is the regularization term with ${\norm*{f}_{\mathcal{H}_k}^2} = \innerprod{f}{f}_{\mathcal H_k}$ and $\lambda \geq 0$.
It can be shown that, in this setting, the supervised learning problem has an analytical solution of the form~\cite{kimeldorf_wahba_mat_stats_1070_repr_thm,scholkopf_book_2001_generalized_repr_thm}
\begin{equation}
    \label{eq:krr_prediction}
    f^\star(x) = \sum_{i=1}^{N_S} w_i k(x, x_i),
\end{equation}
 where the sum only goes over the finitely many training samples and
the weights $w_i$ are uniquely determined by solving the linear system of equations
\begin{equation}
    \label{eq:kernelweights}
    \sum_{j=1}^{N_S}\left(k(x_i, x_j) + \lambda\, \delta_{i,j}\right) w_j  = y_i.
\end{equation}
In the case where $k(x_i, x_j)$ is singular, there is an infinite number of $w$ that satisfy \cref{eq:kernelweights}; The presence of the  infinitesimal regularization term $\lambda$ in this equation ensures that we choose the solution with the minimal norm $\norm*{f}_{\mathcal{H}_k}$.

\begin{figure*}[ht!]
\subfloat{\includegraphics[height=0.35\textwidth]{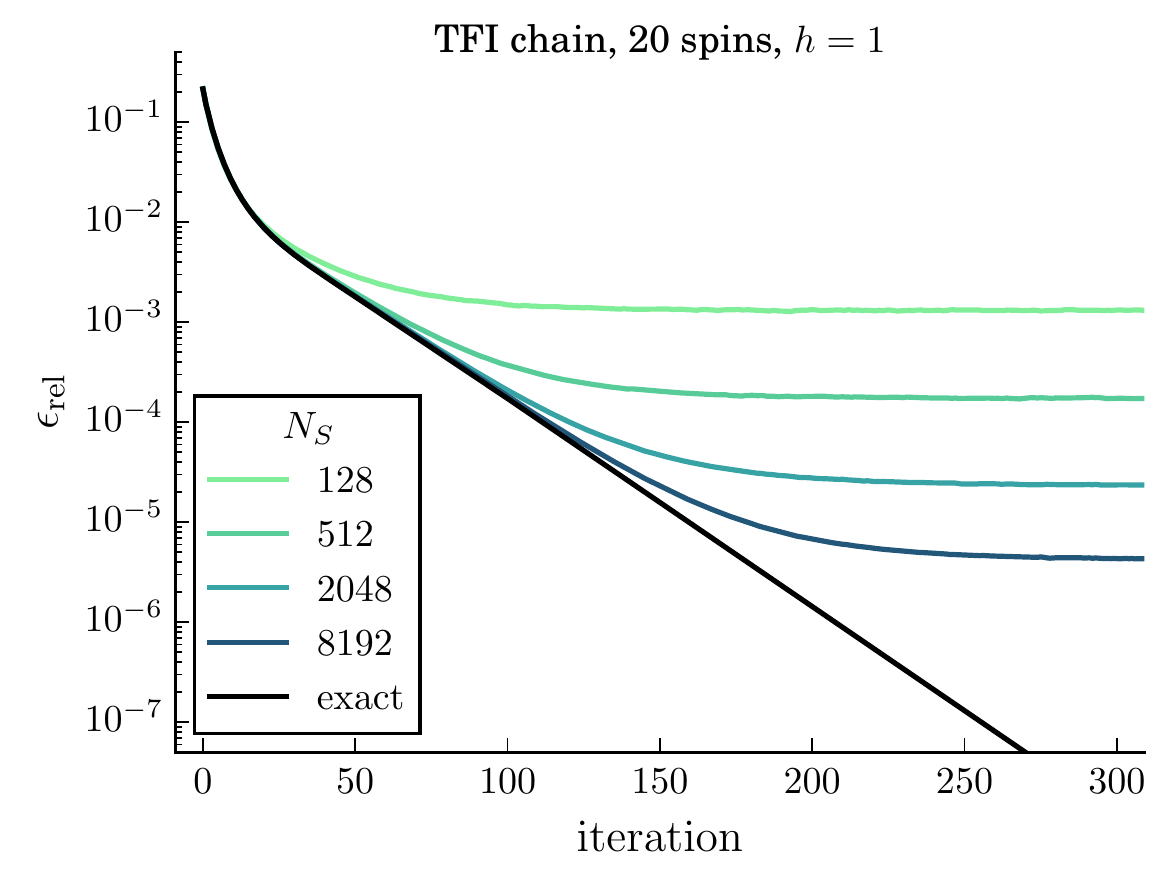}}
\hspace{0.01\textwidth}
\subfloat{\includegraphics[height=0.35\textwidth]{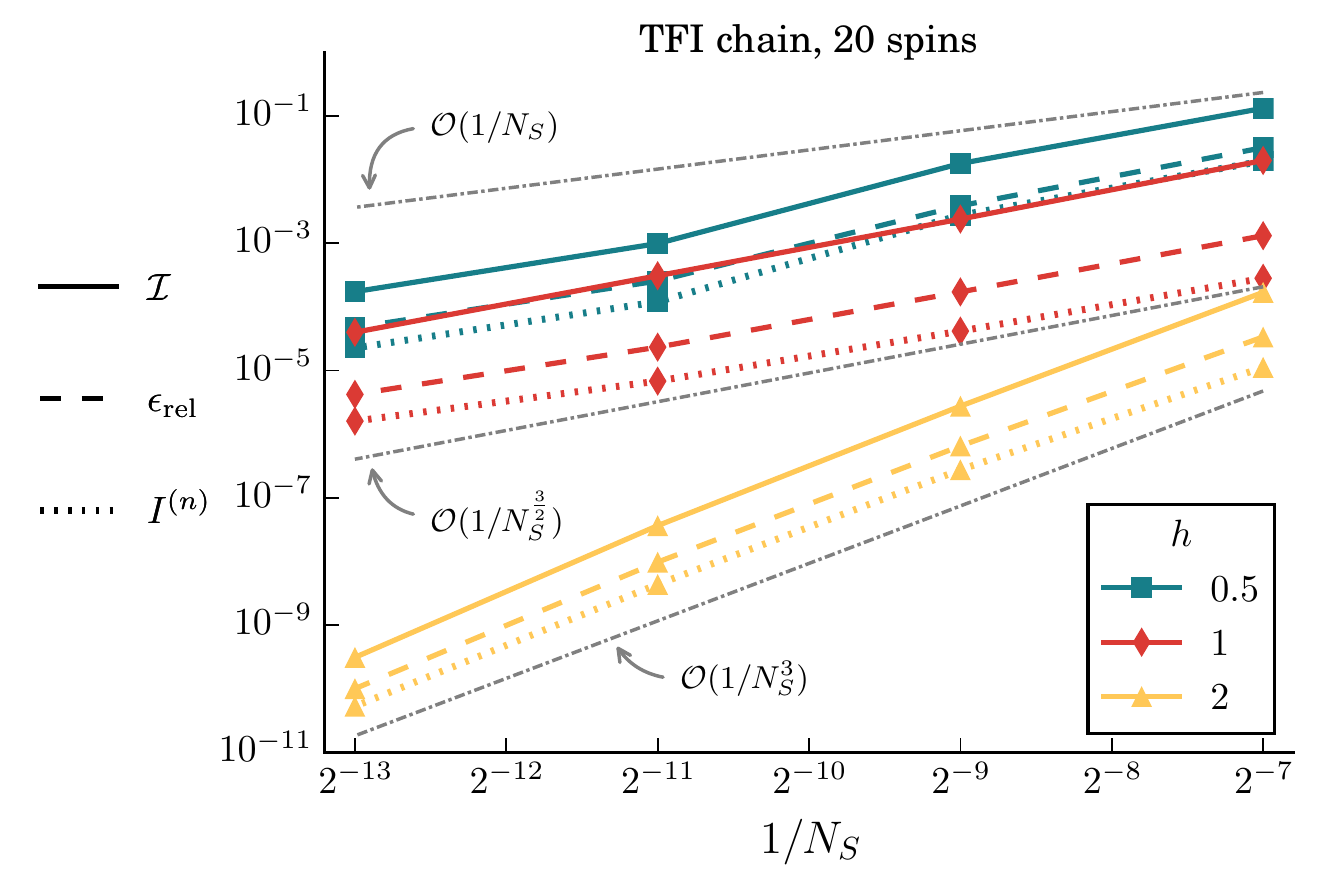}}
\caption{
    Convergence of the Self-learning power method for the TFI model on a one-dimensional chain of $N=20$ spins.
    \textbf{(left panel)}: Relative error of the predicted energy with the true ground state energy as a function of the number of iterations $n$, compared to the power method for $h=1$.
    Starting from an initial uniform superposition state, after a certain number of iterations, a steady state is reached, with an energy that becomes more accurate with increasing data-set size $N_S$, taking the average over 100 runs.
    \textbf{(right panel)}: Final state convergence.
    Plotted are $I^{(n)}$: step infidelity of learning the final state (see \cref{def:stepfid}), $\mathcal I$:  infidelity of the final state with the true ground state (defined in \cref{eq:final_infidelity_bound_epsilon}), and $\epsilon_{\mathrm{rel}}$: relative error of the predicted energy of the final state (defined in \cref{eq:rel_err_bound}) after convergence of the self-learning power method, as a function of the number of samples in the data-set $N_S$. Statistical error bars are smaller than the markers and have been omitted from the plot.
}
\label{fig:fig2}
\end{figure*}
\subsection{Implementation}
\label{sec:slpm-with-krr}
The most straightforward approach would be to learn the amplitudes $\Psi^{(n+1)}$ with functions from the reproducing kernel Hilbert space  $\mathcal{H}_k$ (\cref{eq:reproducing_kernel_hilbert_space}) using kernel ridge regression. However, as is commonly done with NQS, and in a previous work using a different kernel method (Ref. \cite{glielmo_booth_prx_2020_gaussian_process_states}), we use the method to learn the log-amplitudes $\log \Psi^{n+1}(x)$ instead.
The loss function remains the same, but the data-set $D^{(n+1)} = \{(x_i, y_i)\}$ is changed to include log-amplitudes as labels,
\begin{equation}
\begin{aligned}
\label{eq:data-set_log}
    x_i &\sim \Pi(x)\\
    y_i &= \log\mel*{x_i}{\Lambda-\hat H}{\Psi^{(n)}}
\end{aligned}
\end{equation}
and we have to take the exponential to make predictions
\begin{equation}
    \Psi^{n+1}(x) = \exp{\sum_{i=1}^{N_S} w_i k(x, x_i)},
\end{equation}
where the weights $w_i$ are found through \cref{eq:kernelweights} for the modified data-set of \cref{eq:data-set_log}.

We remark that this prediction is different from what one would obtain with relevance-vector regression \cite{tipping_mit_1999_rvm,tipping_jmlr_2001_rvm} used in Refs.~\cite{glielmo_booth_prx_2020_gaussian_process_states,rath_booth_jcp_2020_bayesian_opt_quantum_state}. 

While both methods aim to minimize the mean-squared error on the training dataset (first term in \cref{eq:krr_loss}), they favour different solutions. The KRR finds predictions with small RKHS norm $\norm*{f}_{\mathcal{H}_k}$ due to the regularization term (favouring smooth $\log \Psi$). The relevance-vector regression instead favours solutions which have small weights $w$, due to a zero-mean gaussian prior on them~\footnote{The variance of the gaussian prior is given by hyperparmeters found by a type II maximum likelihood optimization\cite{tipping_pmlr_2003_rvm_hyperparam_opt}, which, in particular, can force some weights to be zero.}.

We make one final approximation for the simulations in this article to reduce the computational cost. We assume that the distribution of the previous state $\Psi^{(n)}$ is sufficiently close to the distribution of the propagated state $(\Lambda-\hat H)\Psi^{(n)}$, and sample from \begin{equation}
    \Pi(x) = \abs{\Psi^{(n)}(x)}^2,
\end{equation} instead of $\abs*{(\Lambda-\hat H)\Psi^{(n)}(x)}^2$, reducing the number of evaluations of $\Psi^{(n)}$ required.

\subsection{Kernel choice and symmetries}
\label{sec:kernel_choice}

The properties of the kernel $k(\cdot, \cdot)$ are fundamental, as they are reflected on the encoded wave-function.
For example, discrete symmetries can be explicitly enforced by constructing a kernel that averages the output over all possible input permutations.

In this article, we consider a symmetric kernel of the form
\begin{equation}
\label{eq:ntk_rbmsymm}
    k(x, y) = \frac{1}{|G|} \sum_{g \in G} \sigma \Big(\frac{1}{L}\sum_{i=1}^{L}{(g\, x)_i\, y_i}\Big),
\end{equation}
where $x,y$ are vectors which encode the basis states, $\sigma(x)=x\arcsin(\gamma x)$ is our choice of non-linear function.
We remark that by taking $\gamma \approx 0.5808$, this kernel corresponds to a symmetrized Restricted Boltzmann Machine (RBM) in the infinite hidden-neuron density limit through the neural tangent kernel theory~\cite{jacot_2018_neurips_neural_tangent_kernel}. Details of this connection are explained in \cref{appendix:kernel_derivation} but are not needed for the discussion.
To contain the computational cost, when simulating lattice systems, we consider the group of all possible translations rather than taking the full space-group of the lattice.
Additionally, spin-inversion ($\mathbb{Z}_2$) symmetry can be enforced by choosing an even non-linear function $\sigma$.

\section{Numerical experiments}
\label{sec:results}

To numerically investigate the viability of the SLPM we benchmark it on the transverse-field Ising model (TFI) with periodic boundary conditions in one and two dimensions, and on the antiferromagnetic Heisenberg model (AFH) on the square lattice.

\subsection{TFI model in one dimension at fixed system size}
The Hamiltonian of the TFI model is
\begin{equation}
    \hat H_{TFI} = \sum_{\langle i,j \rangle} \hat{\sigma}^z_i \hat{\sigma}^z_j \, - h \sum_{i} \hat{\sigma}^x_i,
\end{equation}
where $\hat{\sigma}^{x,y,z}_i$ are Pauli matrices on site $i$,  $\langle i,j \rangle$ iterates over all nearest-neighbor pairs, and $h$ is the strength of the external field in the transverse direction.

We start by considering a 1-dimensional chain of 20 spins with transverse field $h\in\{0.5, 1, 2\}$.
The initial state is always taken to be $\ket{\Psi^{(0)}} \propto \sum_x\ket{x}$, the uniform superposition of all computational basis states, and we fix $\Lambda=1$.

In the left panel of \cref{fig:fig2}, we plot the relative error of the energy with respect to the ground-state value as a function of the number of iterations for $h=1$.
We compare the SLPM for several data set sizes $N_s$ against the exact version, observing a crossover from an initial regime where the effect of the noise is negligible. The SLPM closely matches the exact one to a regime where the noise dominates and the bound given by \cref{thm:thm2.3} prevents further improvements and a steady-state is reached.
As expected, the number of steps at which we observe the crossover depends on the number of samples.

\begin{figure*}[ht!]
\vspace{0.25cm}
\subfloat{\includegraphics[height=0.35\textwidth]{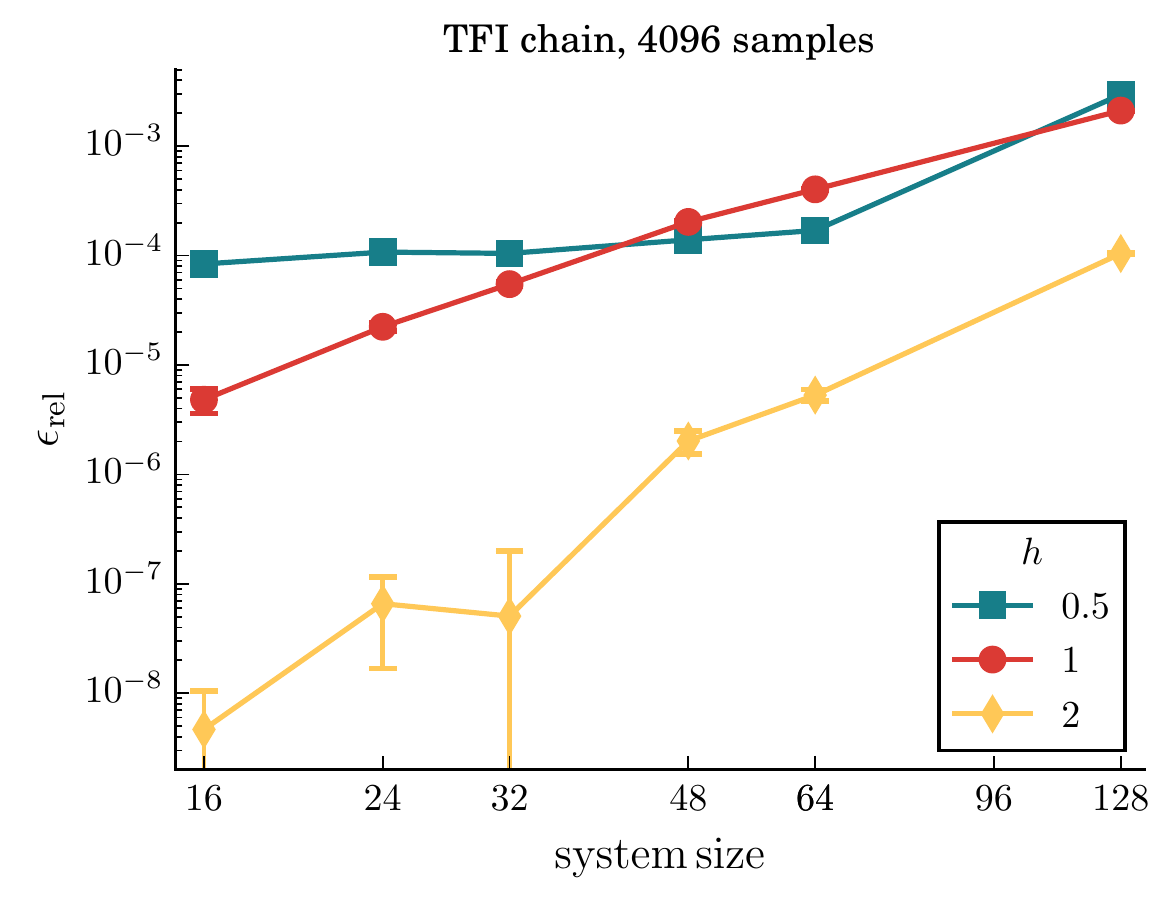}}
\hspace{1cm}
\subfloat{\includegraphics[height=0.35\textwidth]{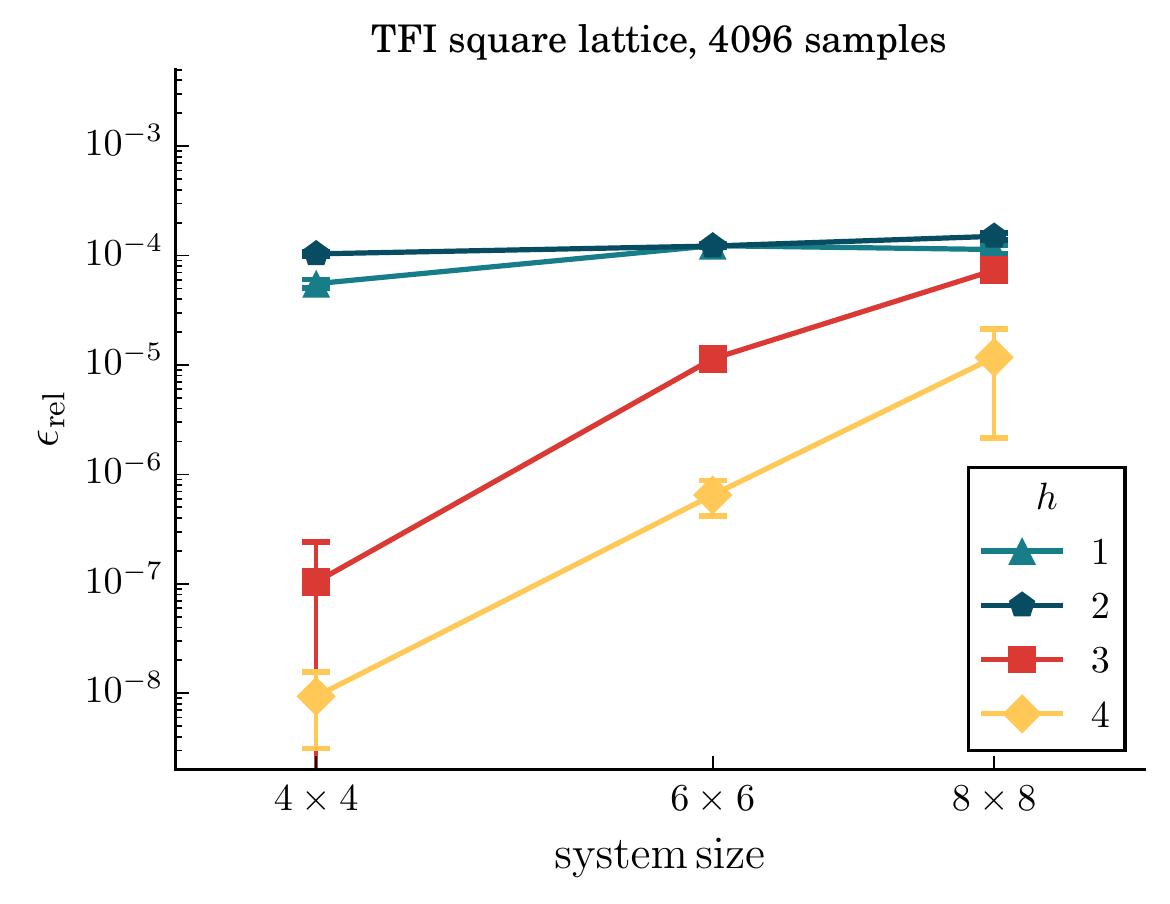}}
\vspace{0.25cm}
\caption{
    Scaling of the SLPM ground-state energy relative error as a function of the system size for 1D \textbf{(left panel)} and 2D \textbf{(right panel)} periodic lattices of the TFI Hamiltonian with varying values of the transverse field $h$. The number of samples is fixed in all simulations at $N_S=4096$ and the energies are estimated by taking $2^{20}$ samples from the final state. The horizontal axes use a logarithmic scale of the total number of spins.
    The reference energies for the relative error are computed analytically for $1$-D systems, with exact diagonalization for $2$-D up to $40$ sites (using the code from Ref.~\cite{westerhout_joss_2021_lattice_symmetries,westerhout_github_2020_spined}) and with Quantum Monte Carlo for larger systems (loop algorithm from the ALPS library~\cite{albuquerque_j_mag_materials_2007_alps1_3,bauer_j_stat_mech_2011_alps2}). They are provided in \cref{appendix:extra_figures}.
}
\label{fig:fig3}
\end{figure*}

\subsection{Numerical verification of the efficient-learning assumption}
In the right panel of \cref{fig:fig2}, we numerically prove the assumption of efficient learning by showing that the step-infidelity at $n=300$ is compatible with a power law $I^{n}\propto N_{S}^{-\alpha}$, where the exponent $\alpha$ depends on the parameters of the Hamiltonian.
In the same figure, we also report that the best relative error follows a similar power law with the same exponent.
Interestingly, we see that the scaling exponent $\alpha$ of the step-infidelity and relative error are degraded for values of $h$ below the critical point ($h=1$ in this case).
This shows that \cref{corr:self_learning_power_method_convergence} is valid and therefore gives further grounding to the theoretical analysis we carried out in \cref{sec:convergence}.
In \cref{fig:fig9} of the appendix we show the same for the TFI in two dimensions and for the AFH in one and two dimensions.

\subsection{TFI model in one and two dimensions and scaling as function of system size}
Continuing, we investigate the scaling of the accuracy of the SLPM at increasing system sizes.
In \cref{fig:fig3} we plot the relative error of the ground-state energy for 1D (left panel) and 2D (right panel) periodic lattices.
The data-set size is kept fixed at $N_S=4096$ for all system sizes.
In both cases, for values of the transverse field $h$ above the critical point\footnote{the critical point of the TFI Hamiltonian in the thermodynamic limit is $h=1$ in $1$-D chains and $h \approx 3.044$ for $2$-D square lattices~\cite{elliott_prb_1970_tfi2d, dejongh_prb_1998_tfi2d_dmrg, rieger_epjb_1999_tfi2d_cluster_algo, bloete_pre_2002_tfi2d, albuquerque_alet_prb_2010_tfi2d}.} we observe a behavior consistent with a power-law dependency of the relative error with the system size.
As the Hilbert-space size is increasing exponentially, this means that a tiny fraction of the Hilbert space is sufficient to compute the ground-state energy accurately in a few hundred steps.
Evidently, at the critical point, the gap of the Hamiltonian becomes smaller and therefore we need to perform more iterations ($M\approx2000$) to converge.

As in the previous simulations, the scaling with the system size degrades for values of the transverse field below the critical point.
This is linked to a \textit{less efficient} supervised learning of the state (in terms of the number of samples) at every step of the SLPM, and is probably related to poor generalization properties of the kernel in this regime.
In principle, we expect that by choosing a different kernel function, it should be possible to improve the learning efficiency and, therefore, the algorithm's overall performance.

\begin{figure*}[ht!]
 \vspace{0.25cm}
\subfloat{\includegraphics[height=0.35\textwidth]{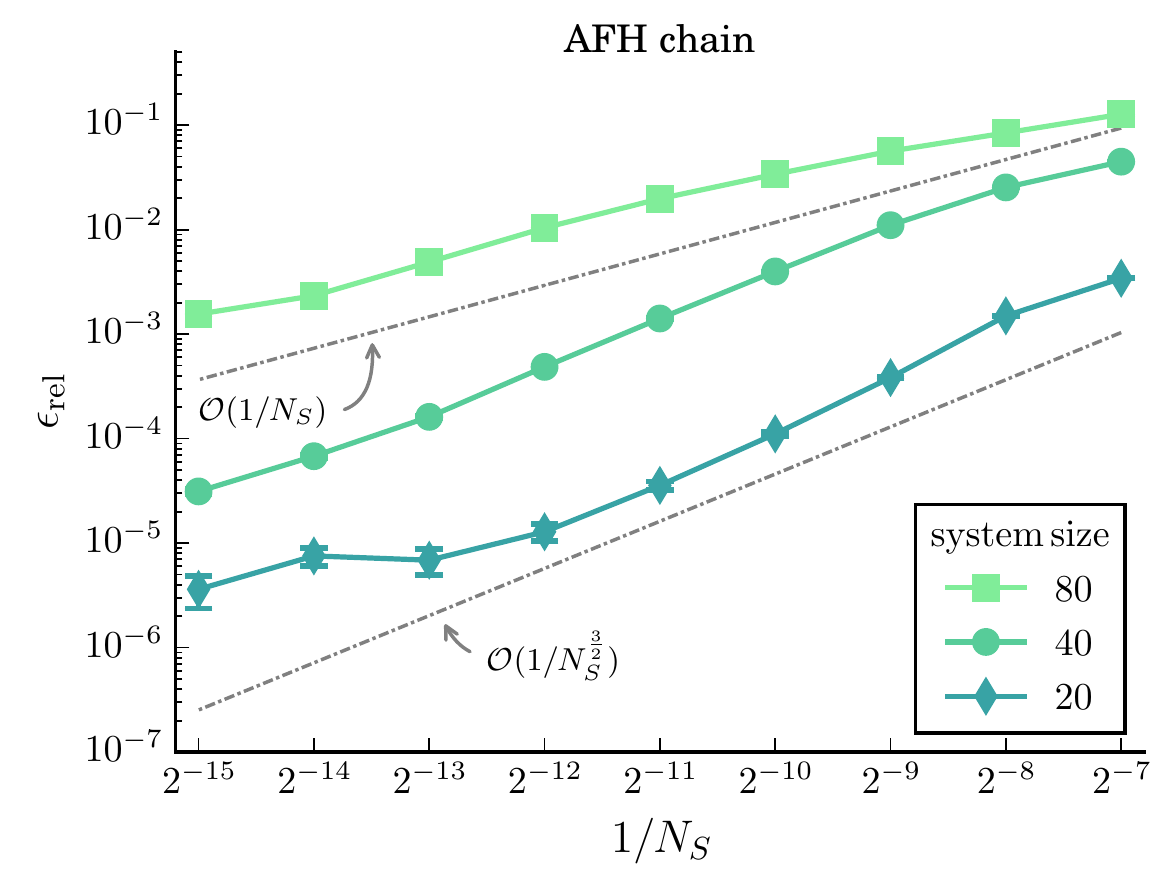}}
\hspace{1cm}
\subfloat{\includegraphics[height=0.35\textwidth]{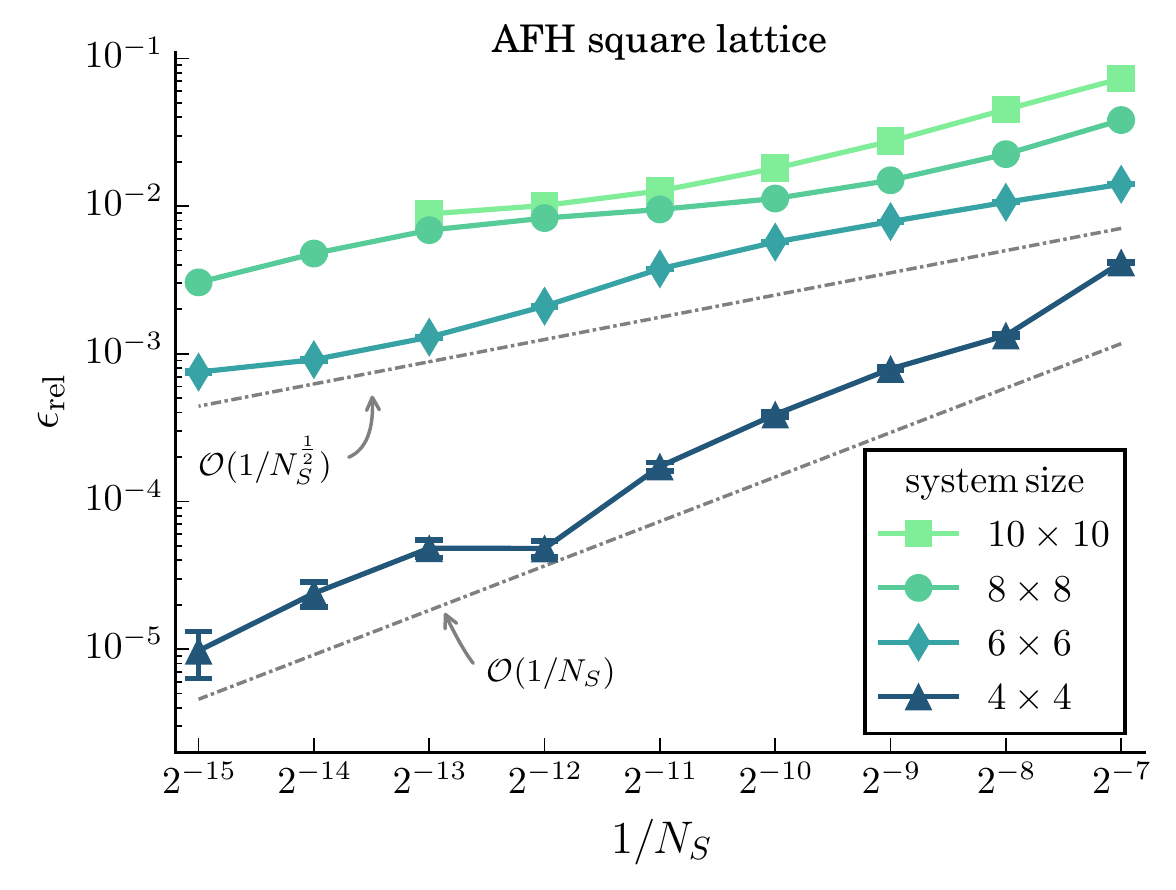}}
 \vspace{0.25cm}
\caption{
    Scaling of the SLPM ground-state energy relative error as a function of the number of samples in the data-set size for 1D \textbf{(left panel)} and 2D \textbf{(right panel)} periodic lattices of the AFH Hamiltonian for different systems sizes.
    Estimates and reference values computed as in \cref{fig:fig3}.
}
\label{fig:fig4}
\end{figure*}

\subsection{AFH model in one and two dimensions}
In addition to the TFI, we also benchmark the SLPM against the antiferromagnetic Heisenberg model, whose Hamiltonian is given by
\begin{equation}
    \hat H_{AFH} = \sum_{\langle i,j \rangle} \hat{\sigma}^x_i \hat{\sigma}^x_j + \hat{\sigma}^y_i \hat{\sigma}^y_j + \hat{\sigma}^z_i \hat{\sigma}^z_j,
\end{equation}
where we assume periodic boundary conditions.
The AFH hamiltonian is gapless in the thermodynamic limit. However, the gap is nonvanishing on finite lattices, and the SLPM can be applied.
The ground state has a well-known sign structure, which can be accounted for by rotating the Hamiltonian according to the Marshall sign rule\cite{marshall_royal_socitey_1955_marshall_sign_rule}.
The SLPM is then used to learn the amplitudes.
To simplify the problem, the ground-state search is constrained to the symmetry sector with zero magnetization by introducing a proper constraint in the sampling step used to generate the data set.
For the simulations of the AFH we fix $\Lambda=0$.
In \cref{fig:fig4}, we show the dependence of the final relative error of the energy as a function of the number of samples in the data set, in the left panel for a one-dimensional chain and in the right panel for two-dimensional square lattices.
Both result in a power-law-like scaling, with an exponent lower than that of the TFI, meaning that supervised learning is less efficient and requires us to use more samples to get a comparable accuracy.

\section{Discussion}
\label{sec:discussion}

In this article, we have presented a kernel-method realization of the SLPM that can be used to find the ground state of gapped quantum hamiltonians by solving a series of quadratic optimization problems.
We have shown that if the supervised learning requires a polynomial number of samples at each step of the power method, a logarithmic number of steps in both the initial overlap and final infidelity of the SLPM is sufficient to reach a ground state infidelity that scales polynomially with the number of samples.
In our numerical experiments, we have considered a relatively simple kernel that is reasonably cheap to evaluate and enforces physical symmetries on the ground state wave-function. 
For the TFI and AFH models in one and two dimensions we have numerically verified that the efficient-learning assumption is valid for kernel ridge regression using this kernel, at least for the small system sizes for which the exact infidelity computation is tractable.
For larger system sizes, a direct computation shows a favorable scaling of the energy relative error in terms of number of samples and as a function of the system size.

Our kernel ridge regression approach is ultimately limited by the number of samples as the computational resources needed to compute and store the kernel matrix scale quadratically with the data-set size, while the solution of the linear system of equations scales cubically.
Therefore, in practice the number of samples is at most of the order of ${10}^5$ due to hardware limitations.
Algorithmic improvements such as re-using information from the matrix decomposition of the previous step when most of the samples remain the same or using iterative solvers enabling parallelization could ease some of these limitations.
Nevertheless, we believe that the primary focus for improvements has to be laid on increasing the efficiency of the supervised learning. This can be done either by developing kernels with superior generalization properties for the problems at hand, 
with kernel methods which do not require to compute the full kernel matrix, like those used in Refs. \cite{glielmo_booth_prx_2020_gaussian_process_states,rath_booth_jcp_2020_bayesian_opt_quantum_state}, or by using methods based on neural networks.

Possible extensions of this work include the application of the SLPM to non-stoquastic Hamiltonians.
This can be achieved by learning the sign structure or the phase of the wave-function in addition to the absolute value of the amplitude.
It might be worth exploring a more general Ansatz using pseudo-kernels~\cite{boloix_tortosa_perez_cruz_ieee_2017_complex_kernels}, where the absolute value and phase of the wave-function amplitude are learned simultaneously.

The code used to run the simulations in this article can be found in Ref. \cite{code_repo}.

\section*{Acknowledgements}
The authors would like to thank George Booth for insightful discussions.
This work was supported by the Swiss National Science Foundation under Grant No. 200021\_200336.

\bibliographystyle{quantum}
\bibliography{main.bib}

\clearpage
\appendix
\begin{center}
\textbf{\large Appendix}
\end{center}

We provide a proof of \cref{thm:thm2.3} and touch on the scaling in the case of gapless Hamiltonians in \cref{appendix:proof_thm_2.3}, discuss the details of the numerical implementation of the sampling and kernel ridge regression in \cref{appendix:implementation_details} and give additional numerical results in \cref{appendix:extra_figures}. Finally in \cref{appendix:kernel_derivation} we highlight the connection between the kernel used in this article and symmetrized Restricted Boltzmann Machine (RBM) in the infinite hidden neuron density limit.

\section{Proof of Theorem \ref{thm:thm2.3} (Convergence of the noisy power method)}
\label{appendix:proof_thm_2.3}

We provide a proof of \cref{thm:thm2.3} presented in the main text, conceptually following the steps of the proof in the supplementary material of Ref.~\cite{hard_neurips_2014_noisy_power_method}.

The main idea of the proof is to show that, if the noise is small enough, at every step the distance between the current state and the ground state decreases with respect to the previous step, where the Fubini-Study metric, given by
\begin{align}
    \label{eq:fubini-study-metric}
    \theta(\psi, \phi)
    \coloneqq \arccos \sqrt{\mathcal F(\phi, \psi)}
    =
    \arccos \left(\frac{
    \abs{\braket*{\phi}{\psi}}
    }{
    \norm{\psi}\norm{\phi}
    }\right)
\end{align}
is used as as the measure of distance.
We remark that the the Fubini-Study metric is a monotonic function of the infidelity.
To simplify notation we define the following short-hand notation for the overlap of the state and noise at step $n$ with the eigenstates of the Hamiltoinian
\begin{equation}
\begin{aligned}
    \Psi^{(n)}_k = \braket*{\Upsilon_k}{\Psi^{(n)}}\\
    \Delta^{(n)}_k = \braket*{\Upsilon_k}{\Delta^{(n)}}.\\
\end{aligned}
\end{equation}

We assume that $\Lambda > \frac{E_1 + E_{\mathrm{max}}}{2}$, and therefore the two largest (also in absolute terms) eigenvalues of $\Lambda-\hat H$ are given by $\sigma_0 = \Lambda-E_0$ and $\sigma_1 = \Lambda -E_1$.
Their difference is is equal to the gap of $\hat H$ since $\sigma_0-\sigma_1 = E_1-E_0 = \delta$.

Assuming normalized eigenstates $\norm{\Upsilon_k}=1$ we have that the cosine of the Fubini-Study distance is given by
\begin{equation}
    \cos \theta(\Psi^{(n)}, \Upsilon_0) = \sqrt{\mathcal  F(\Psi^{(n)}, \Upsilon_0) }
    =\frac{\abs{{\Psi^{(n)}_0}}}{\norm{\Psi^{(n)}}}.
\end{equation}
Then, using basic trigonometric identities we find
\begin{equation}
\tan \theta(\Psi^{(n)}, \Upsilon_0)
= \sqrt{\frac{1-\mathcal  F(\Psi^{(n)}, \Upsilon_0)}{\mathcal  F(\Psi^{(n)}, \Upsilon_0)}}
= \frac{\sqrt{\sum_{k\geq1} \abs{{\Psi^{(n)}_k}}^2}}{\abs{{\Psi^{(n)}_0}}}
\end{equation}
where we used $\sum_{k} \abs{{\Psi^{(n)}_k}}^2 = \norm{\Psi^{(n)}}^2$.

We can show that the distance decreases at every step of the noisy power method under the following assumptions on the noise at each step, in terms of the current state $\Psi^{(n)}$:
\begin{equation}
\label{eq::assumptions_previous}
\begin{aligned}
\abs{\Delta^{(n)}_0}
&\leq \frac{\delta}{4} \norm{\Psi^{(n)}} \cos \theta(\Psi^{(n)}, \Upsilon_0)\\
\norm{\Delta^{(n)}} &\leq  \frac{\delta}{4} \norm{\Psi^{(n)}} \varepsilon
\end{aligned}
\end{equation}
where $\varepsilon < \frac{1}{2}$.

For $\ket*{\Psi^{(n+1)}} = \gamma^{(n)}\left((\Lambda - \hat H) \ket*{\Psi^{(n)}} + \ket*{\Delta^{(n)}}\right)$ the tangent of the distance of the propagated state with the ground state, which is a monotonic function of it, can be bounded as
\begin{equation}
\label{eq:tan_bounds1}
\begin{aligned}
    \tan& \theta(\Psi^{(n+1)}, \Upsilon_0) \\
    &=\frac{\sqrt{\sum_{k\geq1} \abs{
    \mel*{\Upsilon_k}{\Lambda - \hat H}{\Psi^{(n)}} + {\Delta^{(n)}_k}
    }^2}}{\abs{
    \mel*{\Upsilon_0}{\Lambda - \hat H}{\Psi^{(n)}} + {\Delta^{(n)}_0}
    }}\\
    &\leq  \frac{
    \sigma_1 \sqrt{\sum_{k\geq1} \abs{
    \ket*{\Upsilon_k}{\Psi^{(n)}}}} + \sqrt{\sum_{k\geq1} \abs{{\Delta^{(n)}_k}
    }^2}
    }{
    \sigma_0 \abs{{\Psi^{(n)}_0}} - \abs{{\Delta^{(n)}_0}}
    }\\
    &\leq  \frac{
    \sigma_1 \sqrt{\sum_{k\geq1} \abs{
    \ket*{\Upsilon_k}{\Psi^{(n)}}}} + \sqrt{\sum_{k} \abs{{\Delta^{(n)}_k}
    }^2}
    }{
    \sigma_0 \abs{{\Psi^{(n)}_0}} - \abs{{\Delta^{(n)}_0}}
    }\\
    &=  \frac{
    \sigma_1 \tan \theta(\Psi^{(n)}, \Upsilon_0) + \frac{\norm{\Delta^{(n)}}}{\abs{{\Psi^{(n)}_0}}}
    }{
    \sigma_0 - \frac{\abs{{\Delta^{(n)}_0}}}{\abs{{\Psi^{(n)}_0}}}
    }\\
    &=  \frac{
    \sigma_1 \tan \theta(\Psi^{(n)}, \Upsilon_0) + \frac{\norm{\Delta^{(n)}}}{\norm{\Psi^{(n)}}\cos \theta(\Psi^{(n)}, \Upsilon_0)}
    }{
    \sigma_0 - \frac{\abs{{\Delta^{(n)}_0}}}{\norm{\Psi^{(n)}} \cos\theta(\Psi^{(n)}, \Upsilon_0)}
    }\\
\end{aligned}
\end{equation}
where in the first step we applied the triangle inequality and bounded with the largest eigenvalue in the nominator and applied the reverse triangle inequality in the  denominator, assuming that $\sigma_0 \abs{{\Psi^{(n)}_0}} \geq \abs{{\Delta^{(n)}_0}}$.
Under the assumptions of \cref{eq::assumptions_previous} and bounding $\frac{1}{\cos \theta} \leq 1 + \tan \theta$ \footnote{This is equivalent to bounding $\frac{1}{\sqrt {\mathcal F}} \leq 1 + \sqrt{\frac{1-\mathcal  F}{\mathcal  F}}$.}
we have
\begin{equation}
\label{eq:tan_bound2}
\begin{aligned}
    \tan& \theta(\Psi^{(n+1)}, \Upsilon_0) \\
    &\leq
    \frac{
    \sigma_1 \tan \theta(\Psi^{(n)}, \Upsilon_0) + \frac{\delta}{4} \varepsilon (1+\tan \theta(\Psi^{(n)}, \Upsilon_0))
    }{
    \sigma_0 - \frac{\delta}{4}
    }\\
    &=
(1-\frac{\xi }{\sigma_1 + 3 \xi})
    \frac{\sigma_1 +  \varepsilon \xi
    }{\sigma_1 + 2 \xi} \tan \theta(\Psi^{(n)}, \Upsilon_0) +
    \frac{\xi }{\sigma_1 + 3 \xi}  \varepsilon\\
&\leq
\max \left(\frac{\sigma_1 +  \varepsilon \xi
    }{\sigma_1 + 2 \xi} \tan \theta(\Psi^{(n)}, \Upsilon_0), \varepsilon\right)
\end{aligned}
\end{equation}
where we defined $\xi \coloneqq \frac{\delta}{4}$ and split $\frac{1}{\sigma_1 + 3 \xi}
    = (1-\frac{\xi }{\sigma_1 + 3 \xi})
    \frac{1
    }{\sigma_1 + 2 \xi}
    $.
Then, realizing that $\frac{\xi }{\sigma_1 + 3 \xi} \in [0,1]$, we used that the weighed mean of two terms is less than the maximum \footnote{Let $\alpha \in [0,1]$, $x,y\in \mathbb R$, then $\alpha x + (1-\alpha) y \leq \max(x,y)$}.
Splitting again
$\frac{
    1}{\sigma_1 + 2 \xi}
    = (1-\frac{\xi }{\sigma_1 + 2 \xi})
    \frac{1
    }{\sigma_1 + \xi}$
we have
\begin{equation}
\begin{aligned}
    \frac{\sigma_1 +  \varepsilon \xi
    }{\sigma_1 + 2 \xi}
    &= (1-\frac{\xi }{\sigma_1 + 2 \xi})
    \frac{\sigma_1
    }{\sigma_1 + \xi}  + \frac{ \xi
    }{\sigma_1 + 2 \xi} \varepsilon\\
    &\leq \max \left(\frac{\sigma_1
    }{\sigma_1 + \xi}, \varepsilon\right).
\end{aligned}
\end{equation}
Since ${(1+\frac{\xi}{\sigma_1})^4 \geq 1 +4\frac{ \xi }{\sigma _1}}$
we can bound $\frac{\sigma_1
    }{\sigma_1 + \xi} \leq (\frac{\sigma_1
    }{\sigma_1 + 4\xi})^{\frac{1}{4}} = (\frac{\sigma_1
    }{\sigma_0})^{\frac{1}{4}} = (\frac{\Lambda -E_1}{\Lambda -E_0})^{\frac{1}{4}}$,
and find
\begin{equation}
\label{eq:tan_cvg}
    \tan \theta(\Psi^{(n+1)}, \Upsilon_0) \leq \max\Big( \varepsilon, \omega \tan \theta(\Psi^{(n)}, \Upsilon_0)\Big).
\end{equation}
where ${\omega \coloneqq \max\left((\frac{\Lambda -E_1}{\Lambda -E_0})^{\frac{1}{4}}, \varepsilon\right)}$. Therefore under the assumptions of \cref{eq::assumptions_previous}, the noisy power method decreases $\tan \theta(\Psi^{(n)}, \Upsilon_0)$ at every step until at some step $M$ it reaches $\tan \theta(\Psi^{(M)}, \Upsilon_0)=\varepsilon$ and then it stays at that distance.

Finally we show that the following strengthened assumptions, given in terms of the initial distance, imply the assumptions in \cref{eq::assumptions_previous}, and thus the convergence of the method.
\begin{equation}
\label{eq::assumptions_initial}
\begin{aligned}
\abs{\Delta^{(n)}_0}
&\leq \frac{\delta}{5} \norm{\Psi^{(n)}} \cos \theta(\Psi^{(0)}, \Upsilon_0)\\
\norm{\Delta^{(n)}} &\leq  \frac{\delta}{5} \norm{\Psi^{(n)}} \varepsilon
\end{aligned}
\end{equation}
It is clear that $\norm*{\Delta^{(n)}} \leq  \frac{\delta}{5} \norm*{\Psi^{(n)}}$ implies $\norm*{\Delta^{(n)}} \leq  \frac{\delta}{4} \norm*{\Psi^{(n)}}$.
From \cref{{eq:tan_cvg}}, since $\omega \leq 1$, it follows that at every step
${\tan \theta(\Psi^{(n+1)}, \Upsilon_0) \leq \max \left(\varepsilon, \tan \theta(\Psi^{(n)}, \Upsilon_0) \right)}$ which implies
${\tan \theta(\Psi^{(n)}, \Upsilon_0) \leq \max \left( \varepsilon, \tan \theta(\Psi^{(0)}, \Upsilon_0)\right).}$
Then, since $\cos \theta = \sqrt{\frac{1}{1+\tan^2 \theta}} \geq 1 - \frac{\tan^2 \theta}{2}$
for $\varepsilon \leq \frac{1}{2}$ this implies
$\cos \theta(\Psi^{(n)}, \Upsilon_0) \geq \min \big(1 - \frac{\varepsilon^2}{2}, \cos \theta(\Psi^{(0)}, \Upsilon_0)\big) \geq \frac{7}{8} \cos \theta(\Psi^{(0)}, \Upsilon_0)$ and thus $\frac{1}{5} \cos \theta(\Psi^{(0)}, \Upsilon_0)\leq \frac{1}{4}\cos \theta(\Psi^{(n)}, \Upsilon_0)$.

\subsection*{Number of steps}
We derive the bound on the number of steps $M$ needed to reach a state with $\tan \theta \leq \varepsilon$.
As $\omega$ in \cref{{eq:tan_cvg}} is a multiplicative factor it is clear that a logarithmic number of steps $M$ suffices to reach $\varepsilon$.
In order to bound the number of steps it is convenient to first bound $\ln(\frac{1}{\omega})$.
Using the definition of $\omega$ we have that
\begin{equation}
\ln(\frac{1}{\omega}) = \min\left(\ln(\frac{1}{\varepsilon}), \frac{1}{4} \ln(\frac{1}{\frac{\Lambda -E_1}{\Lambda -E_0}})\right)
\end{equation}
Assuming $\varepsilon < \frac{1}{2}$, we bound $\ln(\frac{1}{\varepsilon}) \geq \ln(2)$.
For the other term we use ${\ln(\frac{1}{\frac{\Lambda -E_1}{\Lambda -E_0}}) \geq 1-\frac{\Lambda -E_1}{\Lambda -E_0}}$ and thus
\begin{equation}
    \label{eq:delta_inv_bnd}
    \ln(\frac{1}{\omega}) \geq \min\left(\ln(2), \frac{1-\frac{\Lambda -E_1}{\Lambda -E_0}}{4} \right) = \frac{1-\frac{\Lambda -E_1}{\Lambda -E_0}}{4}.
\end{equation}
Then, recursively applying \cref{eq:tan_cvg} until step $M$ where we assume to reach $\varepsilon$ we set
\begin{equation}
\label{eq:tan_cvg2}
    \tan \theta(\Psi^{(M)}, \Upsilon_0) = \omega^M \tan \theta(\Psi^{(0)}, \Upsilon_0) \overset{!}{=} \varepsilon.
\end{equation}
Taking the log on both sides, solving for $M$ and using \cref{eq:delta_inv_bnd} we find
\begin{equation}
    M \leq \frac{4}{1-\frac{\Lambda -E_1}{\Lambda -E_0}}  \ln(\frac{\tan \theta(\Psi^{(0)}, \Upsilon_0)}{\varepsilon}).
\end{equation}

\subsection*{Scaling for gapless hamiltonians}
\label{appendix:gappless}
We briefly analyze the scaling of the SLPM in the case of a gapless Hamiltonian with a gap which closes polynomially with the inverse system size.

In \cref{thm:thm2.3} it is easy to see that the number of steps needed is inversely proportional to the gap $\delta = E_1-E_0$, since $(1- \frac{\Lambda - E_1}{\Lambda - E_0})^{-1} = \frac{\Lambda -E_0}{\delta}$.
If we assume for a system of size L that, (I) the gap closes as $\delta \propto L^{-\beta}$, (II) we can choose a constant $|\Lambda| \ll E_0$ and (III) the ground state energy scales as $|E_0| \propto L$,  we have that $\frac{\Lambda-E_1}{\Lambda -E_0} = \frac{\delta}{\Lambda -E_0} \propto L^{-(\beta+1)}$ and thus a number of steps proportional to $L^{\beta+1}$ are needed.
We note that the same applies to the power method in absence of noise.

In the case of noise however, we also need a step infidelity which is small enough so that our method can resolve the gap (see \cref{corr:self_learning_power_method_convergence}). Under the assumptions above this would result in a required step infidelity $ 
\propto L^{-2({\beta+1})}$, meaning that we would need in the order of $ {L^{\frac{2(\beta+1)}{\alpha}}}$ samples. Therefore, as long as the gap closes polynomially with the inverse system size, the method is efficient.

\section{Implementation Details}
\label{appendix:implementation_details}

\subsection*{Details of the Monte-Carlo sampling procedure}
\label{appendix:sampling}
We use Markov-Chain Monte Carlo sampling with the Metropolis-Hastings algorithm (see e.g. Ref.~\cite{becca_sorella_book_2017} Sec. 3.9 and references therein) to generate samples $x\sim \abs{\phi}^2$, where $\phi$ is given in terms of log-amplitudes by the kernel ridge regression predictor
\begin{equation}
    \log \phi(x) = \sum_{i} w_i\, k(x, x_i).
\end{equation}
For the TFI model we perform single-spin flip updates, and for the AFH model we propose to exchange two spins at each step, initializing the markov chains with states which have total spin 0.

\setlength{\tabcolsep}{13.3pt}
\begin{table*}[t]
\begin{tabular}{llcccc}
\hline  \hline
 system size \hspace{0.25cm}  & method \hspace{0.5cm} & $h=1.0$ & $h=2.0$ & $h=3.0$ & $h=4.0$  \\ \hline
 $4\times4$ & ED & $-34.010598$ & $-40.190194$ & $-51.448129$ & $-66.223620$  \\ 
 $6\times6$ & ED & $-76.523833$ & $-90.407093$ & $-115.23271$ & $-148.83265$  \\ 
 $8\times8$ & QMC& $-136.043(2)$ & $-160.722(2)$ & $-204.632(3)$ & $-264.569(3)$  \\ 
 $10\times10$ & QMC&  $-212.567(2)$ & $-251.132(2)$ & $-319.615(3)$ & $-413.379(4)$  \\ \hline \hline 
\end{tabular}
\caption{\label{tab:tfi2dref} Reference ground state energies of the the transverse-field Ising model in two dimensions with periodic boundary conditions for several values of $h$. ED computed with SpinED~\cite{westerhout_joss_2021_lattice_symmetries,westerhout_github_2020_spined}, and QMC with Alps~\cite{albuquerque_j_mag_materials_2007_alps1_3,bauer_j_stat_mech_2011_alps2}.}
\end{table*}

\subsection*{Multiple occurences of the same sample}
Monte-Carlo sampling can produce the same sample multiple times.
Formally, including multiple occurrences of the same samples in the data-set reduces the rank of the kernel matrix, which can lead to numerical instabilities when decomposing the matrix, and needs to be accounted for with the regularization.
We note that when using a kernel which is symmetric with respect to a symmetry group $G$, then two samples $x, x^\prime$ belonging to the same orbit, meaning that $g x = x^\prime$ for some $g\in G$, have the same effect, and we consider them to be the "same" as well.

By introducing a weight factor $c_i$ into the loss, we can account for the number of occurences in the loss, and keep only a unique set of samples.
This reduces the cost, as the kernel matrix is smaller since there are fewer samples, while formally keeping the loss invariant, and results in a de-facto sample-dependent regularization as we see in the following.
The modified loss is given by
\begin{equation}
    \mathcal L = \argmin_{w} \sum_i c_i |f(w;\,  x_i) - y_i|^2 + \lambda\, \lVert f \rVert^2,
\end{equation}
and the optimal weights are given by the solution of
\begin{equation}
    \label{eq:kernelweights_sample_dep_reg}
    (k(x_i,x_j) + \lambda\, \frac{\delta_{i,j}}{c_i}) w_j = y_i
\end{equation}
This is very similar to the original system of equations (\cref{eq:kernelweights} in the main text), except that the regularization is now sample-dependent, except in the limit $\lambda \to 0$, where removing repeated samples from the data-set has no effect.
We found that in practice, when the regularization $\lambda$ is already very small, it is sufficient to remove repetitions, while not using the sample-dependent regularization, and therefore adopt this procedure for the simulations in this paper.

\subsection*{Numerical Details of the supervised learning procedure}
The self-learning power method is initialized with a data-set for the uniform superposition state, taking uniform samples from the whole Hilbert space (taking only states with magnetization 0 for the AFH)  and setting all log-amplitudes for the labels to $y = 0$, resulting in a state $\Psi^{0}(x) = 1$. For the kernel \cref{eq:ntk_rbmsymm} we use the non-linearity $\sigma(x)=x\arcsin(\gamma x)$ fixing $\gamma=0.5808$.
Throughout our experiments we keep the regularization fixed at $\lambda={10}^{-8}$, except in rare cases where we get nan's and have to increase it to ${10}^{-7}$.

We solve the linear system of equations for the weights (\cref{eq:kernelweights}) using the Cholesky decomposition of the regularized kernel matrix.
In general we work with un-normalized states. As the operator $\lambda -\hat H$ is not unitary, to avoid underflow/overflow and ensure numerical stability, at every step we subtract the largest log-amplitude present in the data-set from all the labels, effusively normalizing the state so that $\max_{x_i} \abs{\Psi^{(n)}(x_i)} = 1$ for all samples $x_i$ in the data-set.
We wrote the code for the numerical simulations with the jax library \cite{bradbury_github_2018_jax}, using the sampler from netket \cite{vicentini_scipost_phys_code_2022_netket3, carleo_softwarex_2019_netket} and we optionally parallelized it using mpi4jax \cite{haefner_vicentini_joss_2021_mpi4jax}. All of the simulations in this article were run serially on a NVIDIA V100 gpu.

\FloatBarrier
\section{Additional Numerical Experiments}
\label{appendix:extra_figures}

\setlength{\tabcolsep}{3.1pt}

\begin{table}[ht]
\begin{tabular}{lccc}
\hline \hline
system size & ED & QMC & Ref.~\cite{sandvik_prb_1997_2dafh_sse} \\ \hline
$20$ & $-35.617546$ & n/a & n/a \\
$40$ & $-70.986091$ & n/a & n/a \\
$80$ & n/a & $-141.848(2)$ &  n/a\\
\hline
$4\times4$ & $-44.913933$ & n/a & n/a \\
$6\times6$ & $-97.757590$ & n/a & n/a \\
$8\times8$ & n/a & $-172.414(2)$ & $-172.413(2)$\\
$10\times10$ & n/a & $-268.623(3)$ & $-268.620(2)$  \\
\hline \hline
\end{tabular}
\caption{\label{tab:afhref} Reference ground state energies of the the anti-ferromagnetic Heisenberg model on one and two-dimensional periodic lattices. ED computed with SpinED~\cite{westerhout_joss_2021_lattice_symmetries,westerhout_github_2020_spined}, and QMC with Alps~\cite{albuquerque_j_mag_materials_2007_alps1_3,bauer_j_stat_mech_2011_alps2}. For 2D we also report SSE results from Ref.~\cite{sandvik_prb_1997_2dafh_sse} for comparison.}
\end{table}

\begin{figure*}[htb]
\subfloat{\includegraphics[height=0.25\textwidth]{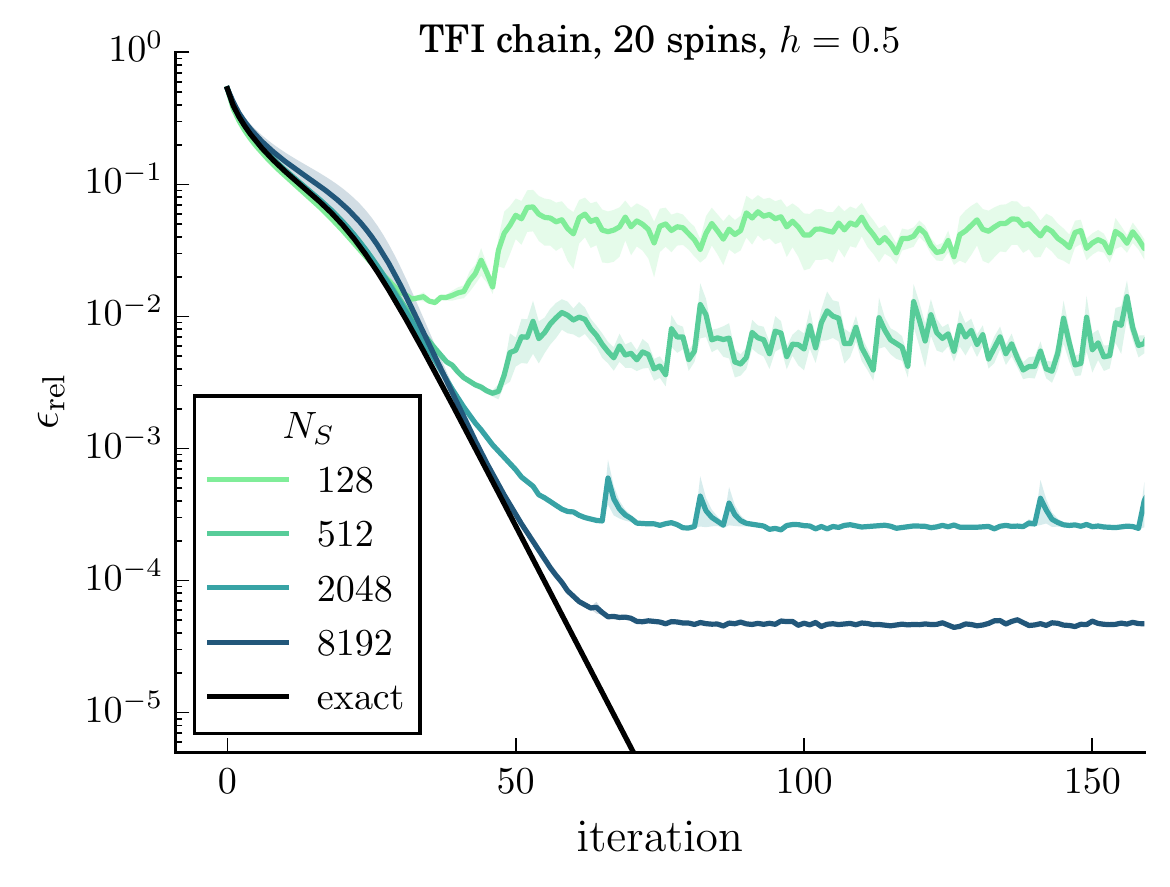}}
\subfloat{\includegraphics[height=0.25\textwidth]{fig/fig2a_5b.pdf}}
\subfloat{\includegraphics[height=0.25\textwidth]{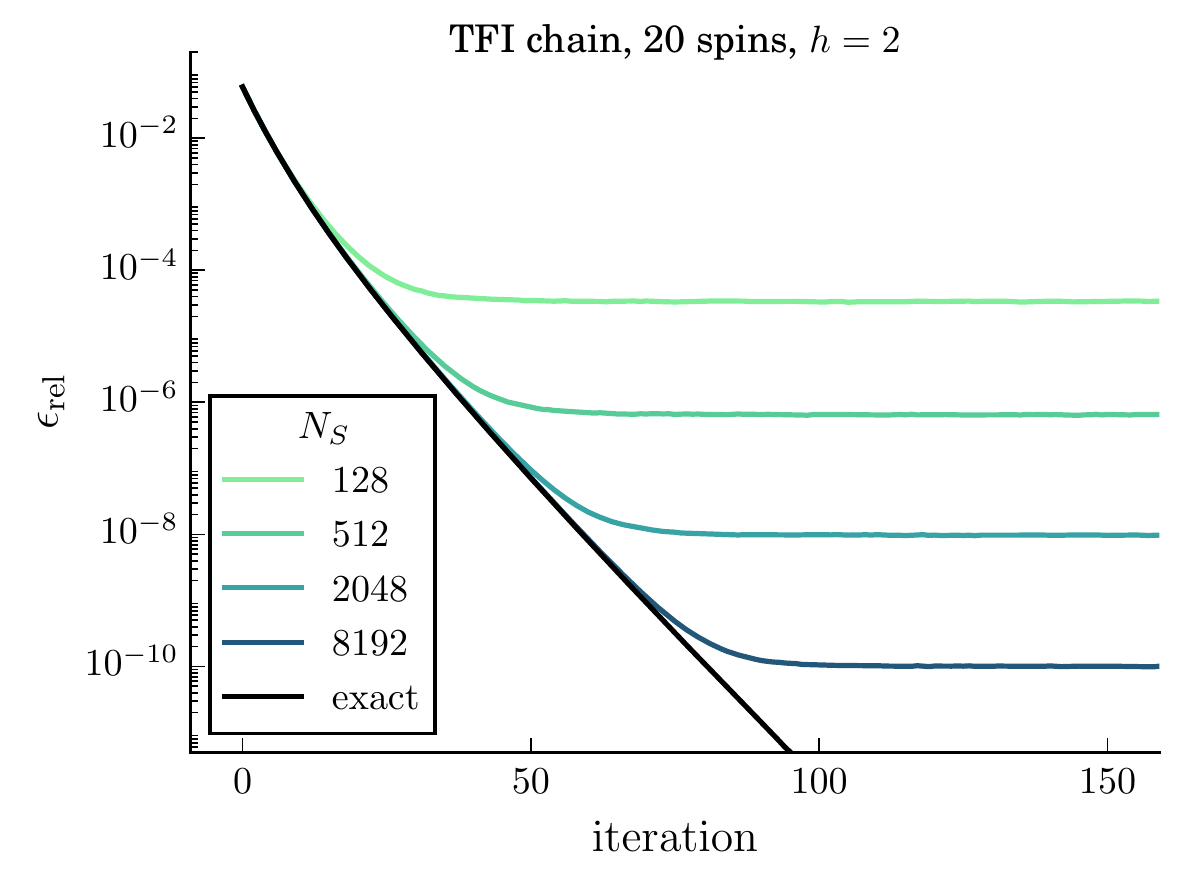}}
\caption{
Convergence of the Self-learning power method for the TFI model on a one-dimensional chain of $N=20$ spins. Relative error of the predicted energy with the true ground state energy as a function of the number of iterations $n$, compared to the power method for $h=0.5$ (left panel), $h=1$ (central panel) and $h=2$ (right panel), taking the average over 100 runs. The central panel is equal to the left panel of \cref{fig:fig2}.
}
\label{fig:fig5}
\end{figure*}

\begin{figure*}[htb]
\subfloat{\includegraphics[height=0.25\textwidth]{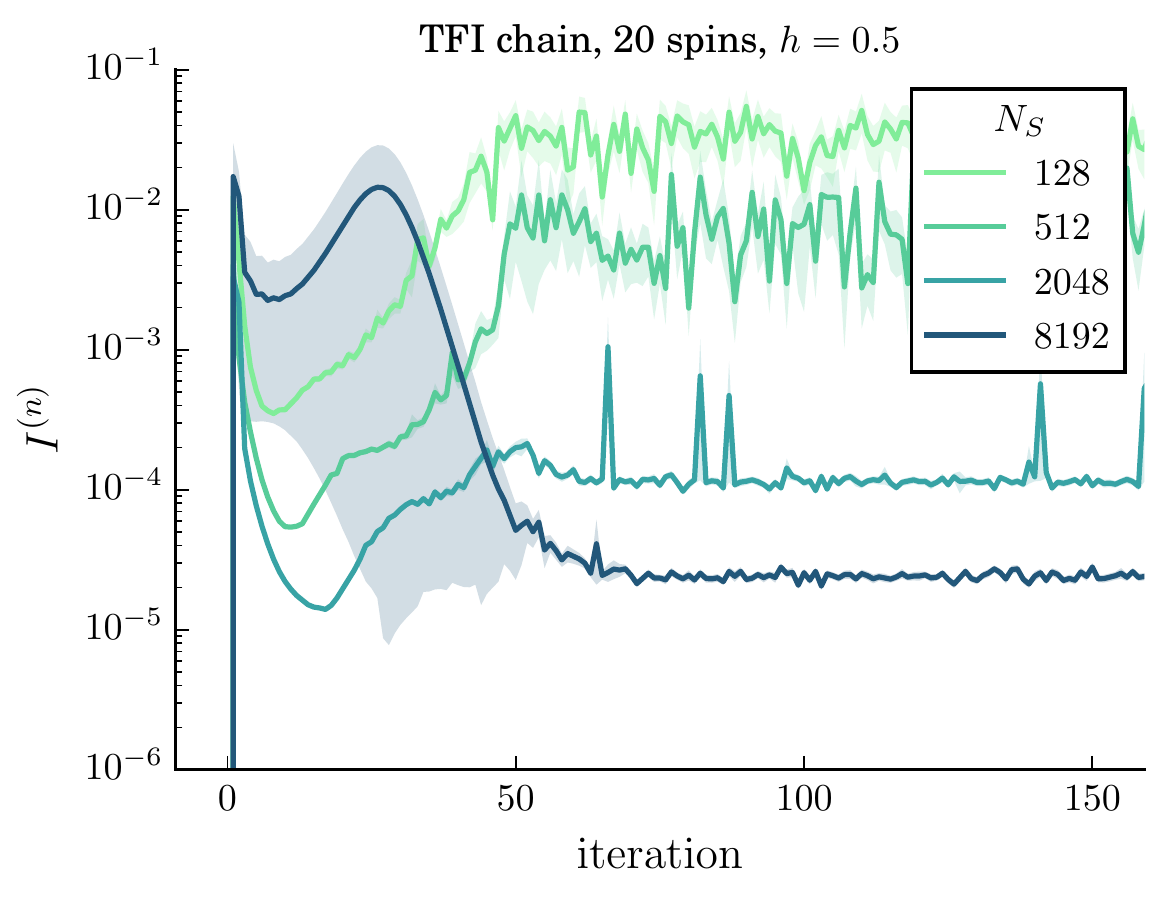}}
\subfloat{\includegraphics[height=0.25\textwidth]{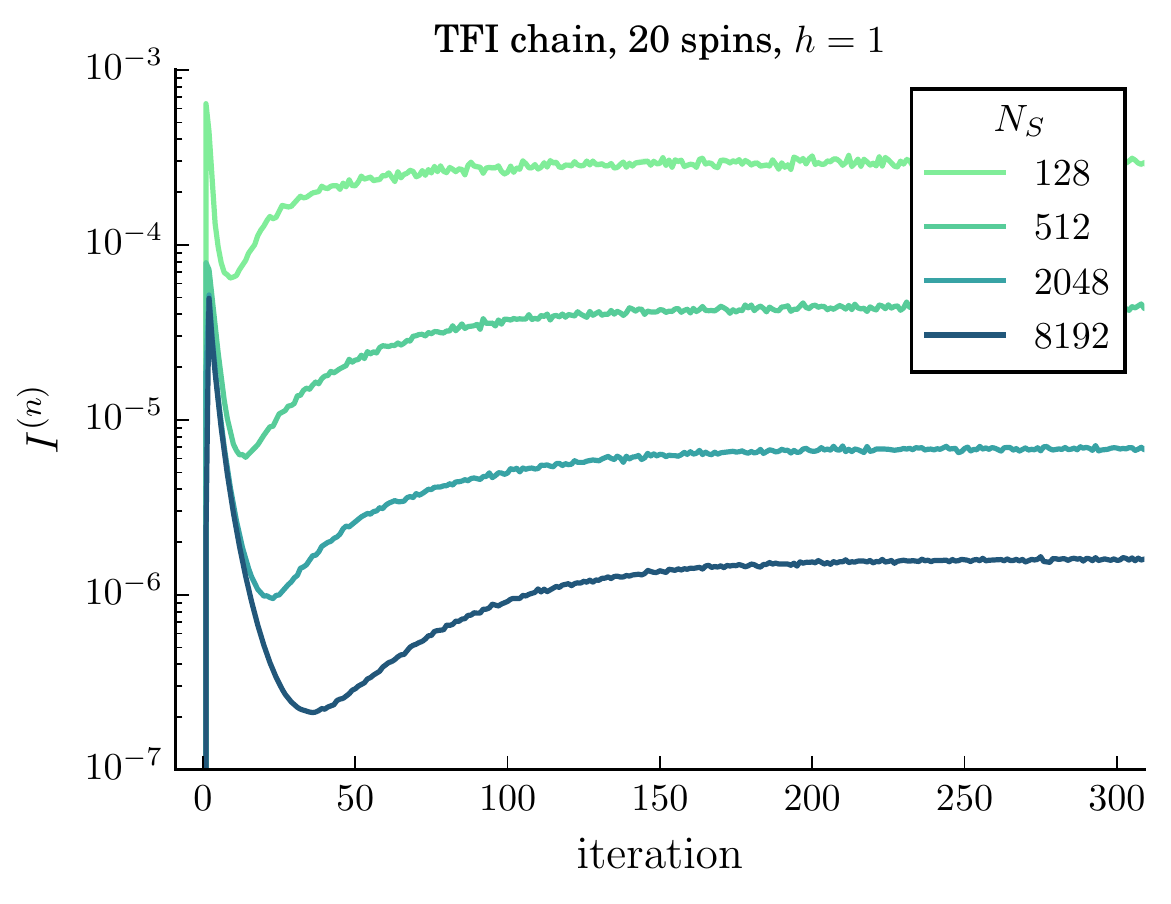}}
\subfloat{\includegraphics[height=0.25\textwidth]{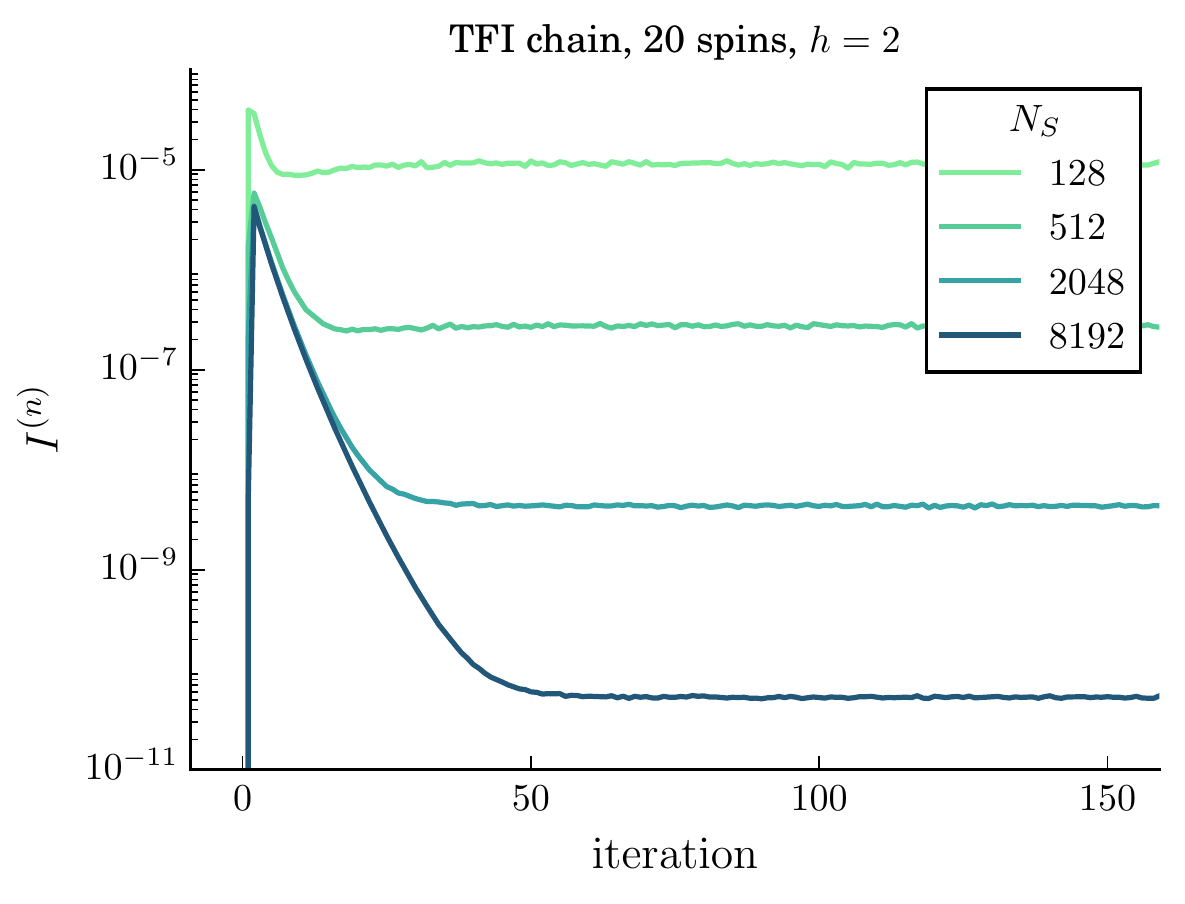}}
\caption{Step infidelity of learning the states visited along the SLPM for the TFI model on a one-dimensional chain of $N=20$ spins. (left panel): $h=0.5$, (central panel): $h=1$, (right panel): $h=2$. Shown are averages over 100 runs.
}
\label{fig:fig6}
\end{figure*}

For completeness, in this appendix we provide the reference energies used for benchmarking purposes and and present several additional numerical experiments that we performed to further support the results in the main text.

In order to benchmark our method we computed reference energies with exact diagonalization (ED) using using the SpinED package~\cite{westerhout_joss_2021_lattice_symmetries,westerhout_github_2020_spined}, and did Quantum Monte Carlo (QMC) simultations using the loop algorithm from the Alps package~\cite{albuquerque_j_mag_materials_2007_alps1_3,bauer_j_stat_mech_2011_alps2}. For the QMC simulations we fixed the inverse temperature at $\beta=1000$ and ran ${10}^5$ thermalization steps followed by ${10}^6$ sweeps.
In \cref{tab:tfi2dref} we provide energies for the TFI model in two dimensions and in \cref{tab:afhref} for the AFH model in one and two dimensions.

\begin{figure*}[htb]
\hspace{-0.25cm}
\subfloat{\includegraphics[height=0.35\textwidth]{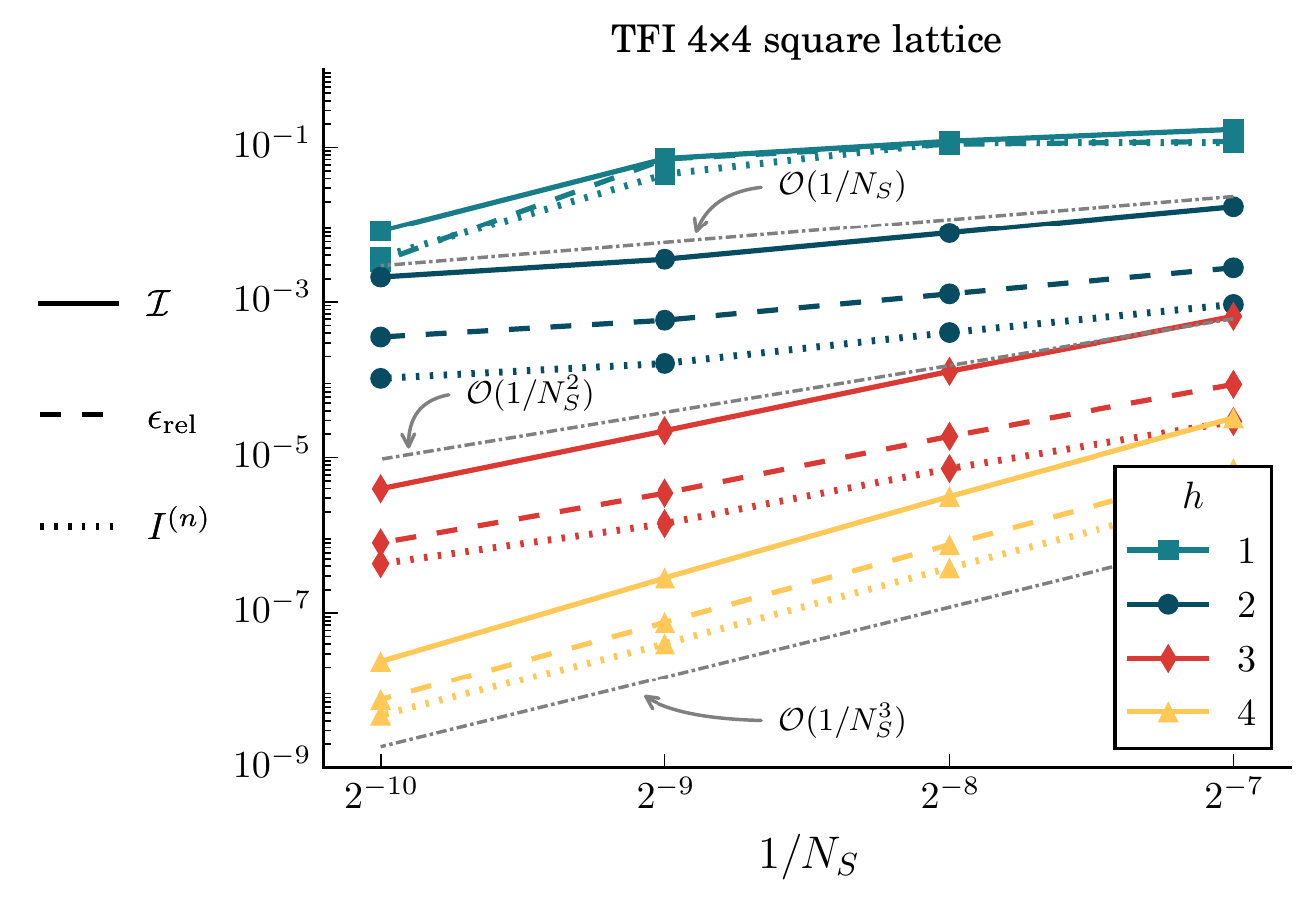}}
\subfloat{\includegraphics[height=0.35\textwidth]{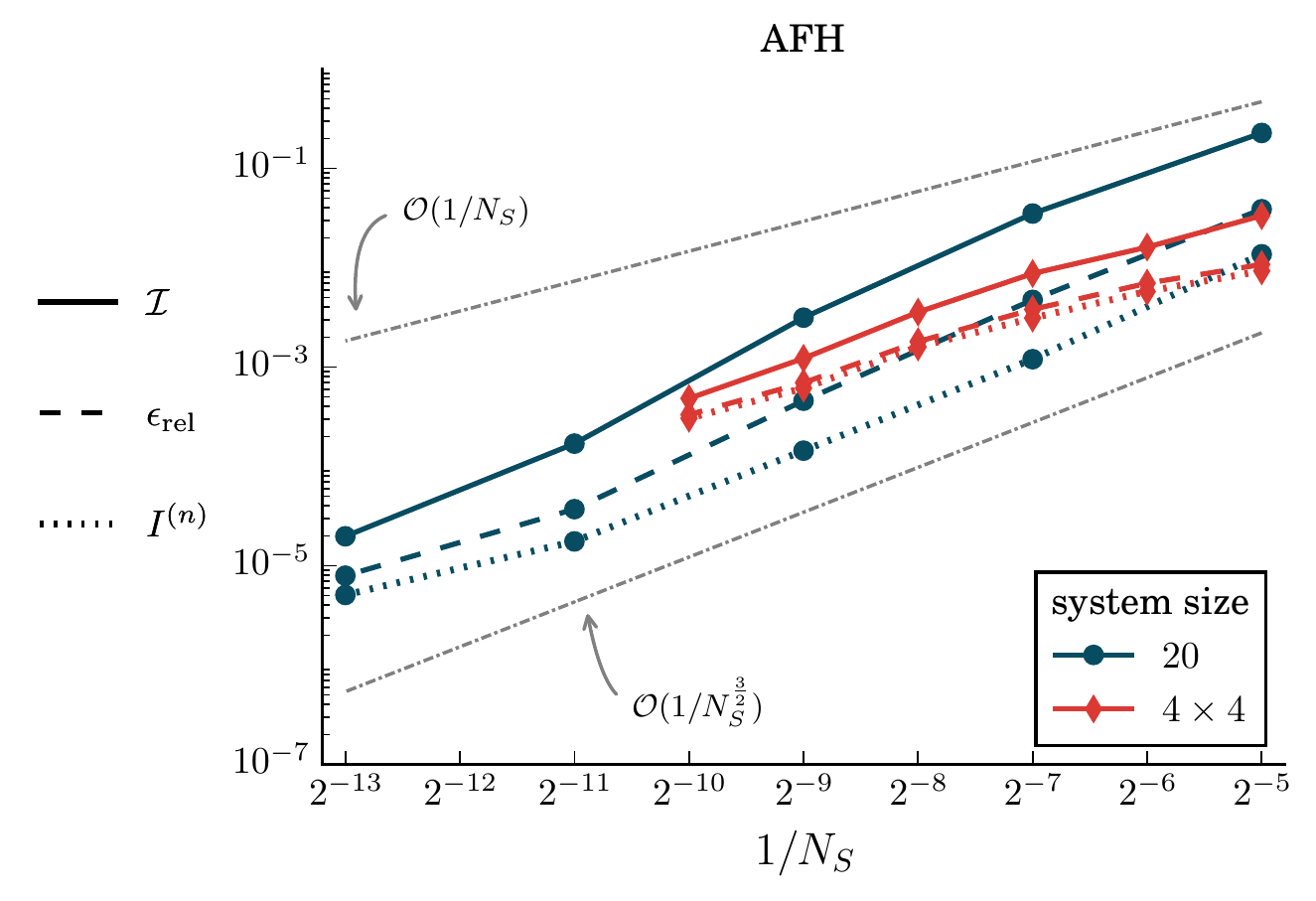}}
\caption{
Final state convergence for the TFI model in 2D (left panel) and for the AFH model in 1D and 2D (right panel), in analogy to the right panel of \cref{fig:fig2} in the main text (TFI in 1D).
 Plotted are $I^{(n)}$: step infidelity of learning the final state (see \cref{def:stepfid}), $\mathcal I$:  infidelity of the final state with the true ground state (defined in \cref{eq:final_infidelity_bound_epsilon}), and $\epsilon_{\mathrm{rel}}$: relative error of the predicted energy of the final state (defined in \cref{eq:rel_err_bound}) after convergence of the self-learning power method, as a function of the number of samples in the data-set $N_S$, taking averages over 100 runs.
}
\label{fig:fig9}
\end{figure*}

\begin{figure*}[htb]
\hspace{-0.25cm}
\hspace{0.88cm}
\subfloat{\includegraphics[height=0.35\textwidth]{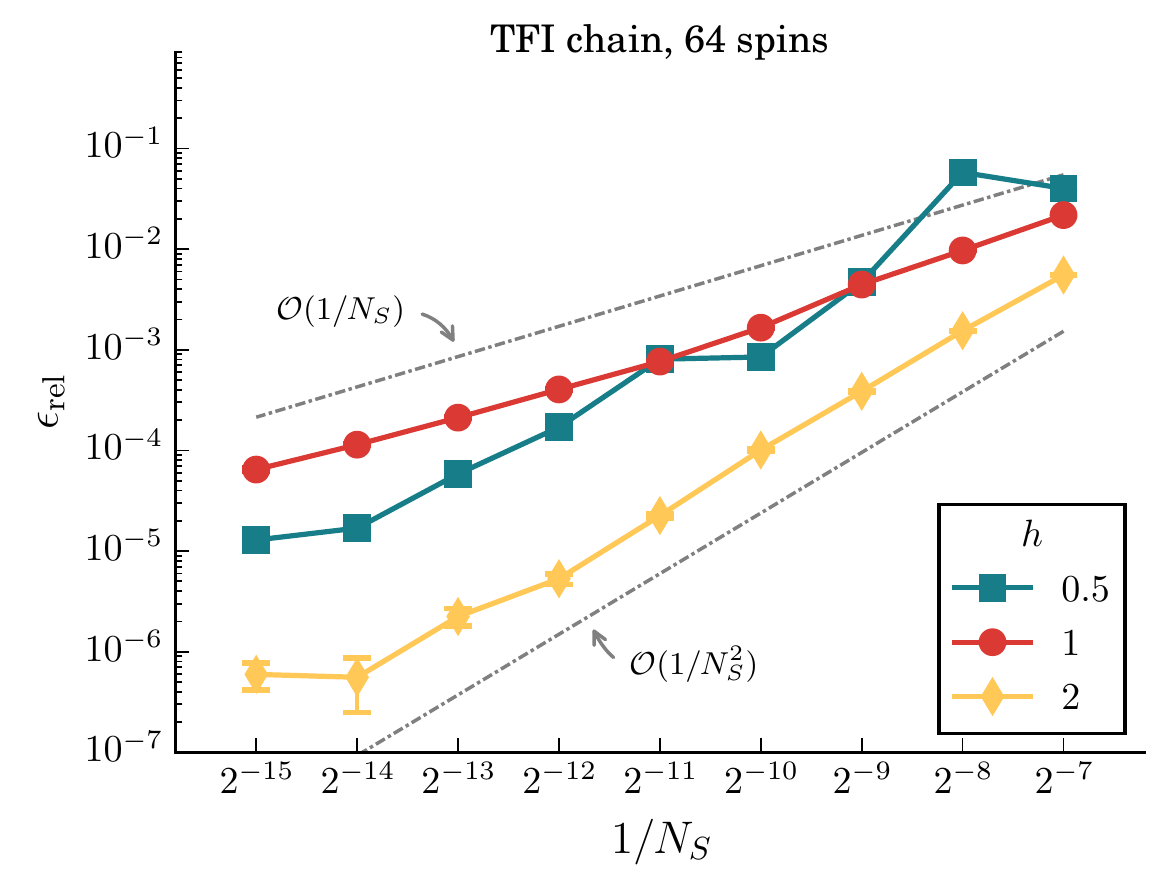}}
\hspace{0.75cm}
\subfloat{\includegraphics[height=0.35\textwidth]{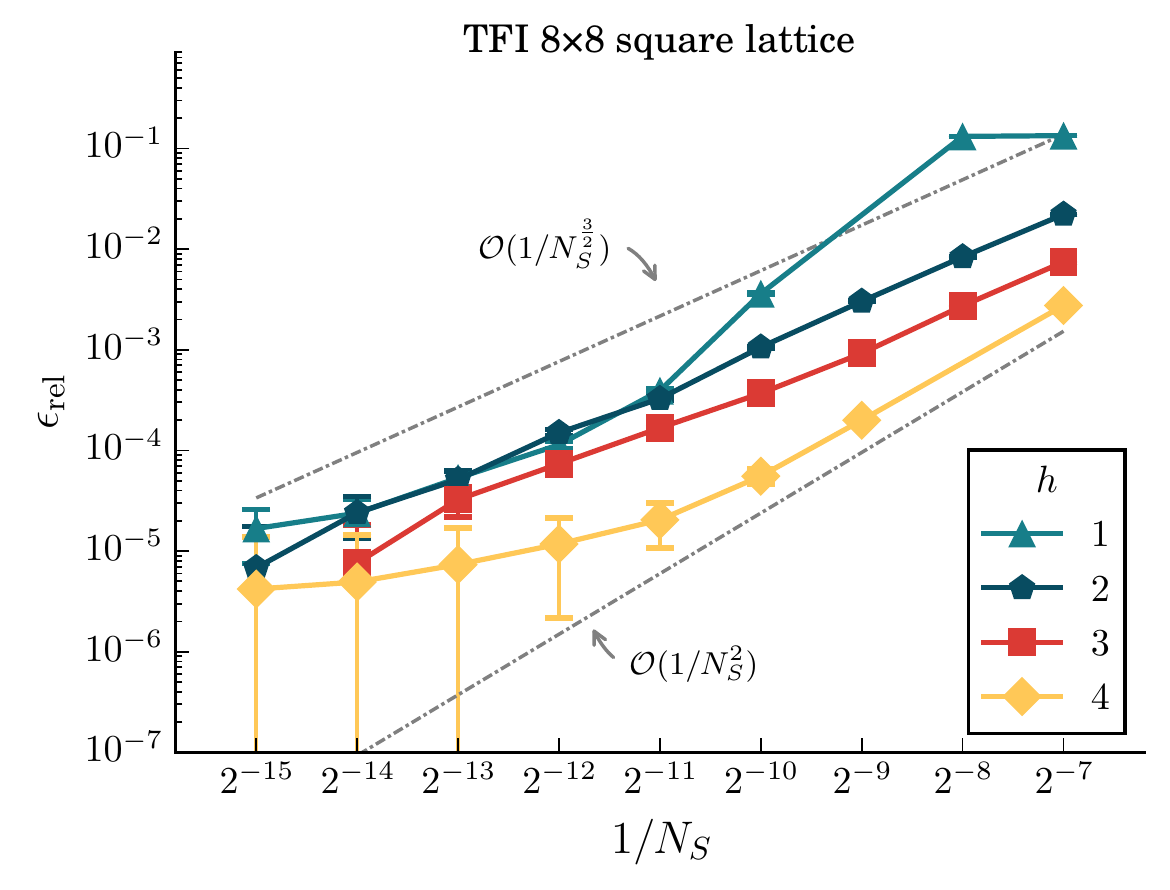}}
\caption{Scaling of the SLPM ground-state energy relative error as a function of the number of samples in the data-set size for a 1D chain of 64 spins \textbf{(left panel)} and 2D $8\times8$ \textbf{(right panel)} periodic lattice of the TFI Hamiltonian.
    Estimates and reference values computed as in \cref{fig:fig3}.
}
\label{fig:fig7}
\end{figure*}

\subsection*{{TFI model in one dimension at fixed system size}}
We start with a few additional results for the 20-spin Ising chain. In \cref{fig:fig5} we provide a plot for the relative error as a function of the number of iterations, for $h=0.5$ and $h=2$ in addition to $h=1$ as already plotted in the left panel of \cref{fig:fig2} in the main text.
We observe comparable behaviour in all three regimes, except for $h=0.5$ when the number of samples is low and fluctuations occur.
This can be explained by the large noise introduced in this case as can be seen by our study of the step infidelity in the following plot.
In \cref{fig:fig6} we plot step infidelity $I^{(n)}$ as a function of the step $n$.
We observe that it is not constant for all the steps of the self-learning power method, but varies before finally leveling off when a steady state is reached, as can be seen by comparing to \cref{fig:fig5}.
In the case of $h=0.5$ and a number of samples $N_S\leq 512$ the step infidelity becomes higher over time, causing the error to increase. We would like to point out that in this case the condition on the step infidelity mentioned in the discussion of \cref{eq:epsilon_relation_step_infidelity} ($\varepsilon < 1/2$) is not satisfied, and therefore \cref{thm:thm2.3} cannot be applied. Nevertheless, by increasing the number of samples we can alleviate these fluctuations.

As pointed out in the main text, it is possible to generalize the theoretical bounds to varying noise (quantified by $\varepsilon$ in \cref{thm:thm2.3}).
One straightforward way to do so is to apply the theorem several times in a row.
Starting from an initial state with possibly exponentially small overlap with the ground state, we fix $\varepsilon = \frac{1}{2}$ and run enough steps until a infidelity of $\varepsilon^2=1/4$ is reached.
In this regime the first assumption of \cref{thm:thm2.3} (\cref{eq:assumption-1}) dominates, as the initial fidelity is much smaller than $\varepsilon$.
Now we are in a state with a finite fidelity of $3/4$, which we use as initial state to apply \cref{thm:thm2.3} again, choosing a smaller value of $\varepsilon$, e.g. as a function of the step infidelity according to \cref{eq:epsilon_relation_step_infidelity}. Now the dominant assumption of \cref{thm:thm2.3} is the second one (\cref{eq:assumption-2}) as the initial fidelity is much larger than $\varepsilon$.
We note that this argument also justifies starting the SLPM with a lower number of samples until a steady state is reached, and increasing afterwards, resulting in a lower compuational cost.

\subsection*{Numerical verification of the efficient-learning
assumption}
In the main text we numerically investigated the efficient-learning assumption for the TFI model on a 1D chain with 20 spins, showing that the step infidelity after a fixed number of steps is compatible with a power-law  $I^{(n)} \propto N_S^{-\alpha}$.
In \cref{fig:fig9} we provide additional evidence that this is also verified for the TFI model in two dimensions and for the AFH model, plotting $I^{(n)}$ as a function of the number of samples after $n=200$ steps of the SLPM.
In the left panel we consider a $4\times4$ square lattice of the TFI model for different values of $h$, and in the right panel a 20 spin chain for the AFH in 1D, and a $4\times4$ square lattice in 2D.
Furthermore we observe that the final Infidelity $\mathcal I$ and relative energy error $\epsilon_\mathrm{rel}$ follow similar power laws with the same exponent as $I^{(n)}$ confirming that \cref{corr:self_learning_power_method_convergence} is valid for the these systems.

\begin{figure*}[htb]
\subfloat{\includegraphics[height=0.35\textwidth]{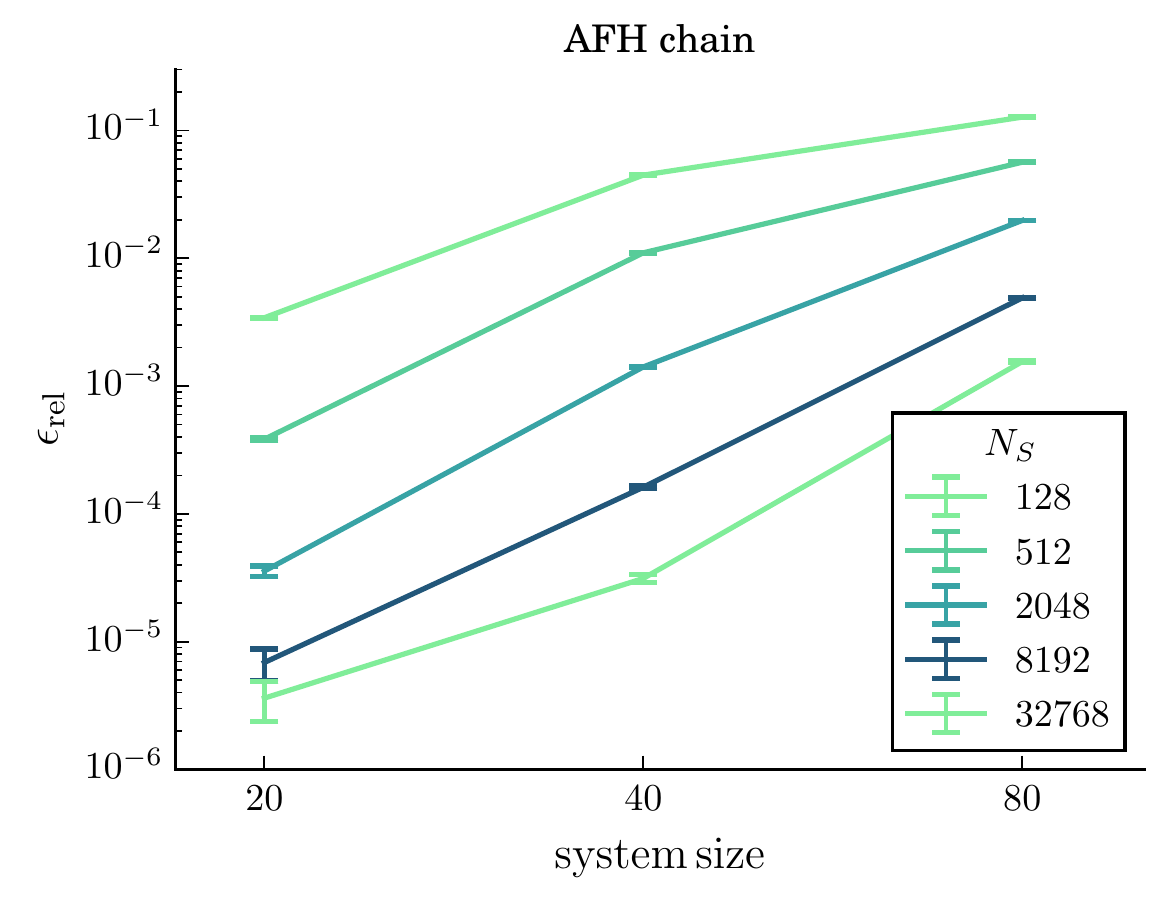}}
\hspace{1cm}
\subfloat{\includegraphics[height=0.35\textwidth]{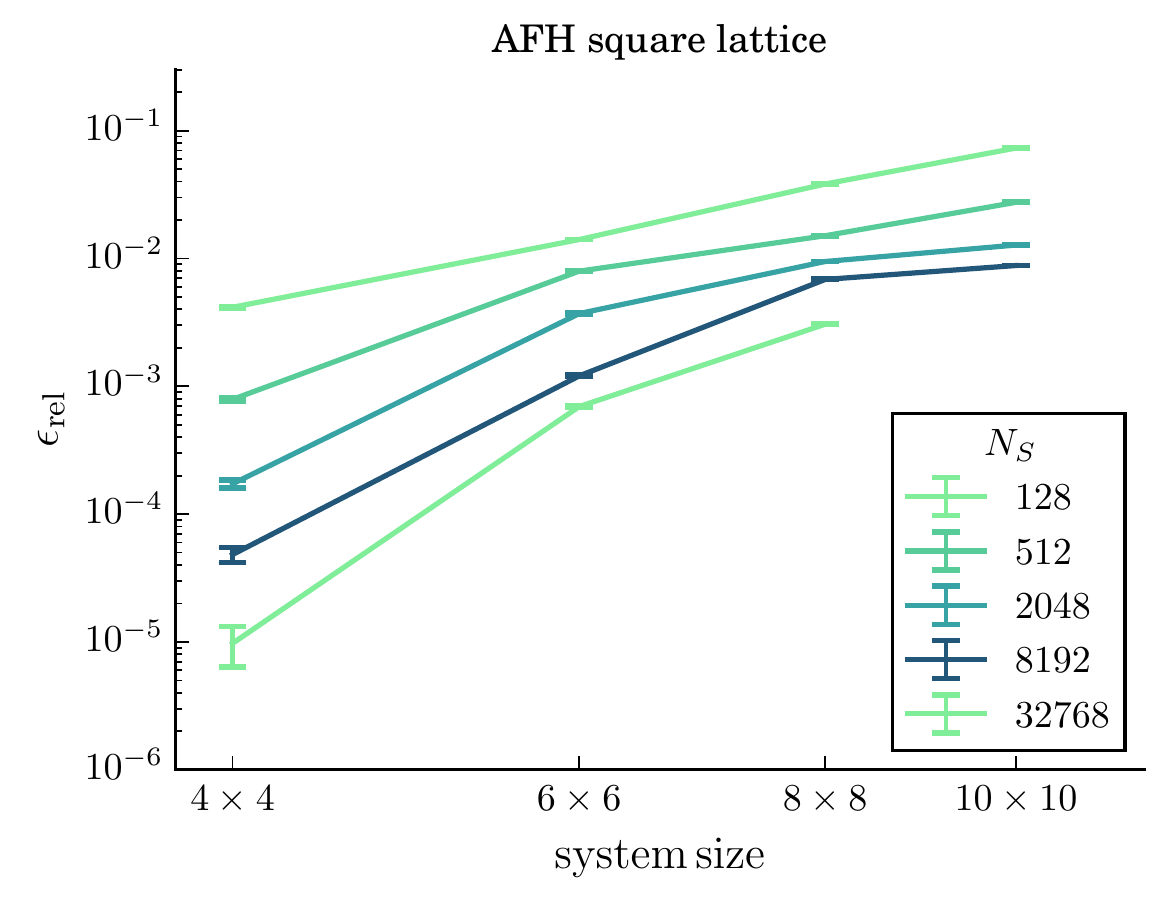}}
\caption{
    Scaling of the SLPM ground-state energy relative error as a function of the system size for 1D \textbf{(left panel)} and 2D \textbf{(right panel)} periodic lattices of the AFH Hamiltonian for different data-set sizes $N_S$. Estimates and reference values computed as in \cref{fig:fig3}.}
    \label{fig:fig8}
\end{figure*}

\subsection*{TFI and AFH model in one and two dimensions}

In the main text we studied the scaling of the SLPM ground-state energy for the TFI model with the system size for a fixed number of samples, and the scaling with the number of samples for the AFH model.
In \cref{fig:fig7,fig:fig8} we provide the respective other plot for the two models.
In \cref{fig:fig7} we study the TFI for 64 spins, in a 1-dimensional chain in the left panel and on a $8\times8$ square lattice on the right panel.
In both cases we find a power law-like scaling of the error with the number of samples, further corroborating our results.
In \cref{fig:fig8} we investigate the scaling of the SLPM error for the AFH in the system size.
For the one-dimensional systems in the left panel we find scaling compatible with a power law, similar to what we found for the TFI in the main text. For the two-dimensional systems we observe that for the largest system and higher number if samples levels off. It might be possible to attribute this to finite-size effects on the smallest system, given that the number of samples becomes of the order of the effective Hilbert space size.

\section{The Kernel of a symmetrized Restricted Boltzmann machine}
\label{appendix:kernel_derivation}

Neural networks, like simple restricted Boltzmann machines (RBM) have been shown to be able to learn the ground states of the systems studied in this article, with an accuracy which increases with the width of the hidden layer, given by the hidden-layer density $\alpha$~\cite{carleo_troyer_science_2017}.
Recently it has been discovered that the training of neural networks is governed by a kernel, the neural tangent kernel (NTK)\cite{jacot_2018_neurips_neural_tangent_kernel}.
It has been shown that in the infinite hidden-layer width limit, on average the prediction of randomly initialized neural networks, which are fully trained with gradiend descent using the least-squares loss, is equal to the prediction of kernel ridge regression using the NTK.
Therefore, this theory formally allows us to study RBM's in the $\alpha \to \infty$ limit using kernel methods.

In particular we are interested in the NTK of a symmetrized RBM, and the effect the symmetrization of the neural network has on the symmetry properties of the kernel.
Let G be a permutation group.
A symmetrized RBM consists of the following layers
\begin{enumerate}[label=(\alph*)]
    \item An initial dense symmetric layer projecting onto $G$
    \item element-wise $\log \cosh$ Nonlinearity
    \item Sum over the features of (a)
    \item Global AvgPool over the elements of $G$
\end{enumerate}
 which produce an invariant model.
We can write it as
\begin{equation}
\label{eq:rbmsymm}
f(\mathbf{x}) =
    \underbrace{\frac{1}{\abs*{G}}\sum_{g \in G}}_{(d)}
    \underbrace{\frac{1}{\sqrt{n_1}}\sum_{i}^{n_1}}_{(c)}
    \underbrace{\sigma}_{(b)}
    (
    \underbrace{\frac{1}{\sqrt{n_0}} \sum_{k=1}^{n_0} (L_g w_i)_k \, \mathbf{x}_k}_{(a)}
    )
\end{equation}
where $\sigma(\cdot) = \log (\cosh(\cdot))$, $w \in \mathbb{R}^{n_1 \times n_0}$, $w_{i,j} \sim \mathcal{N}(0,1)$ is the matrix containing the filters, $\mathbf{x} \in \mathcal X = \{-1,1\}^{n_0}$ is the input of size $n_0$ and $n_1$ is the number of features in the dense symmetric layer of which we take the limit $n_1 \to \infty$. $L_g$ denotes a concrete instance of an operator performing the symmetry operation $g$ on an element of $\mathcal X$, and we are not adding any hidden or visible bias.

The NTK of a neural network $f_\theta(\mathbf{x})$ is defined as
\begin{equation}
    \label{eq:ntk_def_simple}
    \Theta(\mathbf{x},\mathbf{y}) \coloneqq \sum_{p=1}^{P} \partial_{\theta_p}\, f_\theta(\mathbf{x})\, \partial_{\theta_p}\, f_\theta(\mathbf{y})
\end{equation}
where $\theta_p$ are the parameters of the neural network, in our case given by $w$ and $P\to \infty$ when $n_1 \to \infty$.

It can be shown that the NTK of the symmetrized RBM defined above in \cref{eq:rbmsymm} is given by
\begin{equation}
    \Theta(\mathbf x, \mathbf y) \frac{1}{\abs{G}} \sum_{g \in G} \sigma\left(\frac{1}{n_0} (L_g \mathbf{x})^T\, \mathbf{y} \right)
\end{equation}
where $\sigma(x) \coloneqq x \arcsin(\gamma x)$ for $\gamma \approx 0.5808$ which comes from approximating the nonlinearity. We note that in the main text, when presenting this kernel, under a slight abuse of notation we wrote $g$ instead of $L_g$ to simplify the expression.

In the remainder we provide a sketch of the derivation needed to arrive at this simplified expression for the kernel.
We note that the NTK can also be computed with a library such as Ref. \cite{neuraltangents2020}, using circular convolution to get translation symmetry (the full space group is not possible), and with increased computational cost with respect to the kernel presented above, as convolution in twice the number of dimensions as the input needs to be performed.

\subsection*{Derivation of the neural tangent kernel of a symmetrized restricted Boltzmann machine with \textpm 1 input values}
The common way to determine the NTK $\Theta$ of a neural network is to use a recursive procedure computing it layer by layer. To compute $\Theta$ one also needs to compute the covariance $\Sigma$ of the centered multivariate normal distribution describing the output of the neural network over random initialization of the weights, before training.
For a simple feed-forward neural network it has been shown that both $\Theta$ and $\Sigma$ are represented by a scalar kernels times an implicit identity $\otimes I_{n_l}$, as the features of the output of each layer are independent.
For the symmetrized layer of the RBM however, the kernels are no longer scalar, as the same features for different elements of $G$ are correlated. Therefore, we have to work with kernels $\Sigma_{g,g^\prime}$, $\Theta{g,g^\prime}$ which are of size $\abs*{G} \times \abs*{G}$ with $g,g^\prime \in G$. The final scalar NTK is obtained only after the last pooling layer.
We note that it is possible to show, for both the covariance and the NTK for a permutation group $G$, that $\Sigma_{g,g^\prime} = \Sigma_{e,g^\prime g^{-1}}$ and $\Theta_{g,g^\prime} = \Theta_{e,g^\prime g^{-1}}$ holds\footnote{To give a concrete example, in the case of circular translation with stride 1 this means that the kernel matrices are circulant.}, where $e$ is the identity element of $G$. Therefore instead of computing the whole kernel matrix it suffices that we work with a single row.

We apply the recursive procedure to compute the NTK of the symmetrized RBM.

For the first layer (a) we have
\begin{equation}
\label{eq:ntk1}
\begin{aligned}
    \Sigma^{1}_{e,g^\prime} (\mathbf x, \mathbf y)
        &= \frac{1}{n_0} (L_{g^\prime} \mathbf{x})^T\, \mathbf{y} \\
    \Theta^1_{g,g^\prime} (\mathbf x, \mathbf y)
        &= \Sigma^1_{g,g^\prime}(\mathbf x,\mathbf y)= \Sigma^{1}_{e,g^\prime g^{-1}} (\mathbf x, \mathbf y)
\end{aligned}
\end{equation}
where we indicate the layer with the superscript.
After applying the nonlinearity (b) and summing all features (c) we have that the NTK after the second layer is given by
\begin{equation}
 \Theta^{2}_{g,g^\prime}(\mathbf x, \mathbf y) = \dot{\Sigma}^{1}(\mathbf x, \mathbf y)\, \Theta^{1}_{g,g^\prime}(\mathbf x, \mathbf y)
\end{equation}
where
\begin{equation}
    \dot{\Sigma}^{1}(\mathbf x, \mathbf y) = \mathbb{E}_{f \sim \mathcal{N}(0, \Sigma^{1})}[\dot{\phi}(f(\mathbf{x}))\dot{\phi}(f(\mathbf{y}))]
\end{equation}
and  $\dot\phi(\cdot) = \tanh(\cdot)$ is the derivative of $\log \cosh$.

We apply the pooling (d) to get the scalar NTK for the output of the symmetrized RBM, pooling over the whole group $G$:
\begin{equation}
\begin{aligned}
    \Theta^{3}(\mathbf x, \mathbf y)
    &= \frac{1}{\abs{G}^2}\sum_{k \in g G} \sum_{k^\prime \in g^\prime G} \Theta^{2}_{k,k^\prime}(\mathbf x, \mathbf y)\\
    &= \frac{1}{\abs{G}^2} \sum_{k \in G} \sum_{k^\prime \in G} \Theta^{2}_{k,k^\prime}(\mathbf x, \mathbf y)
\end{aligned}
\end{equation}
where we chose $g,g^\prime\in G$ arbitrarly, as $\forall g \in G: gG = G $.

We combine the individual components to find the equation for the NTK of the symmetrized RBM, and simplify to obtain the NTK of the output of the final layer:
\begin{equation}
\label{eq:ntk3}
\begin{aligned}
     \Theta^{3}(\mathbf x, \mathbf y)
     &= \frac{1}{\abs{G}^2} \sum_{k \in G} \sum_{k^\prime \in G} \dot{\Sigma}^{1}(\mathbf x, \mathbf y) \, \Sigma^{1}_{e,k^\prime k^{-1}} (\mathbf x, \mathbf y) \\
     &= \frac{1}{\abs{G}} \sum_{k \in G} \dot{\Sigma}^{1}(\mathbf x, \mathbf y) \, \Sigma^{1}_{e,k} (\mathbf x, \mathbf y)
\end{aligned}
\end{equation}
What is left is to find an expression for $\dot{\Sigma}^{1}(\mathbf x, \mathbf y) = \mathbb{E}_{f \sim \mathcal{N}(0, \Sigma^{1})}[\dot{\phi}(f(\mathbf{x}))\dot{\phi}(f(\mathbf{y}))$, which requires evaluating a two-dimensional gaussian integral.
We are not aware of any analytical solutions for our choice of non-linearity, and opt to approximate the non-linearity with one for which analytical solutions are known.
We can approximate
\begin{equation}
    \frac{d}{dx} \log \cosh(x) = \tanh(x) \approx \erf(\alpha x)
\end{equation}
where $\alpha \approx 0.8324$ (which can be found numerically by minimizing the L2 error in a suitable interval centered around 0).
Then noticing that for all inputs $\mathbf{x} \in \{-1, 1\}^{n_0}$ it holds that $\Sigma^{1}_{g,g}(\mathbf{x}, \mathbf{x}) = \frac{\norm{\mathbf{x}}^2}{n_0} \equiv 1 $, and using Eqs. 10 and 11 of Ref.~\cite{wiliams_adv_neural_info_proc_sys_1996_inf_wide_nn_erf_nonlinearity} we find
\begin{equation}
\begin{aligned}
    \dot{\Sigma}^{1}(\mathbf x, \mathbf y) = \frac{2}{\pi} \arcsin \big(
    \underbrace{\frac{2 \alpha^2}{1 + 2 \alpha^2}}_{\approx 0.5808 \eqqcolon \gamma}
    \Sigma^{1}_{g,g^\prime}(\mathbf{x}, \mathbf{y})\big) \\
\end{aligned}
\end{equation}

Finally we obtain the expression for the kernel presented in the main text (\cref{sec:kernel_choice})
\begin{equation}
\begin{aligned}
     \Theta^{3}(\mathbf x, \mathbf y)
    &= \frac{1}{\abs{G}} \sum_{k \in G} \dot{\Sigma}^{1}(\mathbf x, \mathbf y)\, \Sigma^{1}_{e,k} (\mathbf x, \mathbf y) \\
     &\approx \frac{2}{\pi}\frac{1}{\abs{G}} \sum_{k \in G} \arcsin(\gamma \Sigma^{1}_{e,k}) \Sigma^{1}_{e,k} \\
     &\propto \frac{1}{\abs{G}} \sum_{k \in G} \sigma(\Sigma^{1}_{e,k})
     = \frac{1}{\abs{G}} \sum_{g \in G} \sigma\left(\frac{1}{n_0} (L_g \mathbf{x})^T\, \mathbf{y} \right)
\end{aligned}
\end{equation}
where we substituted $\sigma(x) \coloneqq x \arcsin(\gamma x)$.

\end{document}